\documentclass[a4paper,12pt]{article}
\title{Crosscap states in $\mathcal{N}=2$ Liouville theory}
\author{Yu Nakayama}
\usepackage{amsmath}
\usepackage{amssymb}
\usepackage{graphicx}
\def\drawbox#1#2{\hrule height#2pt
        \hbox{\vrule width#2pt height#1pt \kern#1pt
              \vrule width#2pt}
              \hrule height#2pt}

\def\Fund#1#2{\vcenter{\vbox{\drawbox{#1}{#2}}}}
\def\Asym#1#2{\vcenter{\vbox{\drawbox{#1}{#2}
              \kern-#2pt       
              \drawbox{#1}{#2}}}}

\def\funda{\Fund{6.5}{0.4}}

\def\symm{\funda\kern-0.4pt\funda}

\allowdisplaybreaks[0]

\newcommand{\NS}{\mbox{NS}}
\newcommand{\tNS}{\widetilde{\mbox{NS}}}
\newcommand{\R}{\mbox{R}}

\newcommand{\sNS}{\msc{NS}}
\newcommand{\stNS}{\widetilde{\msc{NS}}}
\newcommand{\sR}{\msc{R}}

\newcommand{\msc}[1]{\mbox{\scriptsize #1}}

\newcommand{\bz}{\mbox{{\bf Z}}}

\newcommand{\bsz}{\msc{{\bf Z}}}

\newcommand{\cN}{{\cal N}}

\newcommand{\cQ}{{\cal Q}}

\newcommand{\sectiono}[1]{\section{#1}\setcounter{equation}{0}}

\begin{document}

\begin{titlepage}
\thispagestyle{empty}
\begin{flushright}
UT-04-26\\
hep-th/0409039\\
September, 2004 
\end{flushright}

\vskip 1.5 cm

\begin{center}
\noindent{\textbf{\LARGE{Crosscap states in $\mathcal{N}=2$ Liouville theory \\\vspace{0.5cm}
}}} 
\vskip 1.5cm
\noindent{\large{Yu Nakayama}\footnote{E-mail: nakayama@hep-th.phys.s.u-tokyo.ac.jp}}\\ 
\vspace{1cm}
\noindent{\small{\textit{Department of Physics, Faculty of Science, University of 
Tokyo}} \\ \vspace{2mm}
\small{\textit{Hongo 7-3-1, Bunkyo-ku, Tokyo 113-0033, Japan}}}
\end{center}
\vspace{1cm}
\begin{abstract}
We construct crosscap states in the $\cN = 2$ Liouville theory from the modular bootstrap method. We verify our results by comparing it with the calculation from the minisuperspace approximation and by checking the consistency with the conformal bootstrap equation. Various overlaps with other known branes are studied. We further discuss the topological nature of the discrete terms in the crosscap wavefunction and their connection with the Landau-Ginzburg approach in a nontrivial dilaton background.  We find that it can be mapped to the Landau-Ginzburg theory with a negative power superpotential by a simple change of variables, extending the known duality to the open string sector. Possible applications to the two-dimensional noncritical string theories and supersymmetric orientifolds in the higher dimension are also discussed.

\end{abstract}

\end{titlepage}

\tableofcontents
\newpage

\sectiono{Introduction}\label{sec:1}
The study of the Liouville theory has regained much interest recently (for a review, see \cite{Teschner:2001rv,Nakayama:2004vk}). One of the main motivations is to study the world sheet description of the two or less dimensional noncritical string theory whose dual description is the exactly solvable matrix model. With the various developments in the nonperturbative aspects of the string theory, the exactly solvable matrix model has again become an arena of the nonperturbative physics from the holographic viewpoint \cite{McGreevy:2003kb,Takayanagi:2003sm,Douglas:2003up}.  

However, as is well-known, the bosonic or $\cN=1$ Liouville theory has a famous $c_{matter} (\hat{c}_{matter}) = 1$ barrier. Thus, the application to the higher dimensional string theory is not so obvious. On the other hand, the $\cN =2$ extended version of the super Liouville theory \cite{Distler:1989nt,Kutasov:1990ua} is liberated from such a constraint because the non-renormalization of the world sheet cosmological constant term eliminates the $\hat{c}_{matter} = 1$ barrier. For this reason, the noncritical string theory based on the $\cN=2$ Liouville theory is not restricted to the lower dimension, and one may formulate the matrix model dual of this theory, which might have a higher dimensional application (some attempts include \cite{McGreevy:2003dn,Verlinde:2004gt}). 

This feature distinguishes $\cN=2$ Liouville theory from $\cN=0,1$ Liouville theory, and enables us to use it as a ``noncompact" inner space of the superstring theory. One of the most famous examples is the CHS (or the near horizon limit of the NS5 brane) background \cite{Callan:1991at}. Likewise, combining $\cN =2$ Liouville theory(ies) with other compact CFT(s), we can construct various exactly solvable CFTs for the string compactification. Typically, the $\cN=2$ Liouville direction is identified with a noncompact direction and the cosmological constant term is related to the deformation of the singularity. Thus, the study of the $\cN=2$ Liouville theory is also relevant for the study of the singularity of the {\it string} geometry such as ADE singularities of ALE space.\footnote{We should emphasize the terminology ``{\it string} singularity" here. In the singular limit, we expect that the massless BPS soliton emerges, and the world sheet description breaks down.}

In this paper, we would like to study the properties of the orientifold plane in the $\cN =2$ Liouville theory, namely the $\cN =2$ Liouville theory on unoriented Riemann surfaces. {}From the space-time point of view, the orientifold plane is another soliton in the superstring theory besides the D-brane, and it becomes a source of the R-R potential much like the D-brane if it emerges in the superstring theory as a BPS soliton. {}From the world sheet point of view, it corresponds to the unoriented Riemann surfaces while the D-brane corresponds to the Riemann surfaces with boundaries. Equivalently, any unoriented Riemann surfaces can be constructed by attaching crosscaps to the oriented Riemann surfaces. Thus, the construction of the crosscap states is one of the central issues regarding the world sheet description of the orientifold plane.

The motivation to study orientifold planes (crosscap states) in this particular noncompact CFT is as follows. In the general string compactification, orientifold planes play a special role. One property is that they become a source of the negative R-R charge while preserving the space-time SUSY. This enables us to cancel the tadpole from the D-branes in the compact direction, which is one of the strongest consistency conditions of the string compactification with D-branes.

 In the noncompact case such as the singular Calabi-Yau spaces represented by $\cN=2$ Liouville theory, the tadpole cancellation is not necessarily required. Nevertheless, even in the noncompact theory the introduction of the orientifold plane yields richer space-time physics such as more general gauge groups ($SO/Sp$ group) and more exotic chiral matter contents. Indeed, the introduction of the orientifold plane has become one of the standard methods to study the dynamics of these supersymmetric gauge theories involving interesting physics such as the chiral matter contents and/or dynamical SUSY breaking to name a few. The noncompactness of the target space corresponds to decoupling of the gravity, which, at the same time, gives the possibility of the dual gravity description inspired by the AdS-CFT correspondence.

The key steps to understand these solitons such as branes and orientifolds in terms of the nontrivial noncompact CFT is the discovery of the FZZT brane \cite{Fateev:2000ik,Teschner:2000md} and the ZZ brane \cite{Zamolodchikov:2001ah} in the bosonic Liouville theory, where the conformal bootstrap method and the modular bootstrap method have been introduced. These methods have been applied to the boundary states of the $\cN = 1$ Liouville theory in \cite{Fukuda:2002bv,Ahn:2002ev}. By using the same technique, the crosscap state in the bosonic string theory has been constructed in \cite{Hikida:2002bt}, and in \cite{Nakayama:2003ep} these results have been applied to the tadpole cancellation problem of the two-dimensional noncritical string theory. The crosscap states in the $\cN =1$ Liouville theory have been discussed in \cite{Gomis:2003vi} and \cite{Bergman:2003yp} in the context of the $\hat{c}=1$ unoriented string theories and their matrix model duals.

On the $\cN =2$ Liouville theory, the branes whose open spectrum is unitary were first classified in \cite{Eguchi:2003ik} by using the modular bootstrap method. This result can be checked by the constraint from the conformal bootstrap method as in \cite{Ahn:2003tt,Ahn:2004qb,Hosomichi:2004ph}. In \cite{Ahn:2004qb}, the ZZ branes in the  $\cN =2$ Liouville theory are also introduced. In this paper, we utilize the results of these works and apply the modular bootstrap method to the M\"obius strip amplitudes in order to obtain the crosscap states in the $\cN=2$ Liouville theory by solving the sewing constraint \cite{Lewellen:1991tb,Fioravanti:1993hf} which corresponds to the Cardy condition \cite{Cardy:1989ir} of the boundary states.

We should note that $\cN = 2$ Liouville theory with a specific compactification is believed to be dual to the $SL(2,\mathbf{R})/U(1)$ supercoset model. This has a proof using the mirror symmetry \cite{Hori:2001ax}. The boundary states of the branes in the $SL(2,\mathbf{R})/U(1)$ model were studied in \cite{Ribault:2003ss,Israel:2004jt,Fotopoulos:2004ut} and compared to the corresponding branes in the $\cN = 2$ Liouville theory. It would be an interesting problem to construct crosscap states in this supercoset theory and to investigate the correspondence between the crosscap states we will construct in this paper.

The organization of the paper is as follows. In section \ref{sec:2}, we study the modular transformation properties of the $\Omega$ inserted characters of the $\cN = 2$ Liouville theory. In section \ref{sec:3}, we derive the crosscap states in the $\cN =2$ Liouville theory from the modular bootstrap method. In the following subsection, we compare it with the semiclassical expectation from the minisuperspace approximation and check its consistency. We study the conformal bootstrap equation and show that it is satisfied by the crosscap states constructed from the modular bootstrap method. We also evaluate the overlaps with other branes and the Klein bottle amplitudes and discuss their physical meanings. In section \ref{sec:top} we point out the topological nature of the discrete terms and the connection to the Landau-Ginzburg approach. In section \ref{sec:5}, we discuss possible applications of our results in the context of two dimensional noncritical strings, supersymmetric compactifications and the orientifold planes in the NS5 brane background. We have two appendices. In appendix \ref{sec:A}, we collect our conventions and useful formulae. In appendix \ref{sec:B}, we present the complete derivation of the modular transformation of the $\Omega$ inserted character including the discrete terms.

\sectiono{$\Omega$ Inserted Characters in $\mathcal{N} =2$ Liouville Theory}\label{sec:2}

First of all, let us introduce the bulk $\cN =2 $ Liouville theory.
The action is given in the superfield formalism by\footnote{The most of the conventions including the superfield are taken from \cite{Nakayama:2004vk}, but one change of the notation is we use $\cQ = b^{-1}$ instead of $b$. The central charge becomes $\hat{c} \equiv c/3 = 1+\cQ^2$ in this convention. Note $\hat{c}$ is denoted by $\tilde{c}$ in \cite{Nakayama:2004vk}. We always set $\alpha'=2$.}

\begin{equation}
S_0 = \frac{1}{2\pi}\int d^2z d^4\theta S\bar S \ ,
\end{equation}
where $S$ is a chiral superfield whose bosonic part we will denote as $S = \phi + iY$. We have a background charge $\cQ$ for $\phi$. In this paper, we mainly focus on the case where $Y$ is noncompact.

The chiral Liouville potential is given by
\begin{align}
S_+ &= \mu \int d^2z d^2\theta e^{\frac{S}{\cQ}} \cr
S_- &= \bar{\mu} \int d^2z d^2\bar{\theta}e^{\frac{\bar{S}}{\cQ}} \ . \label{chil}
\end{align}

The dual non-chiral (Kahler) potential is also possible:
\begin{equation}
S_{nc} = \tilde{\mu} \int d^2z d^4\theta e^{\frac{\cQ}{2}(S+\bar{S})} \ . \label{dual}
\end{equation}
These two possibilities of the marginal interaction term are proposed to be dual to each other \cite{Ahn:2002sx}. Indeed the latter can be regarded as the screening charge of the $SL(2,{\bf R})/U(1)$ supercoset model \cite{Nakayama:2004vk}, which is the mirror dual of the $\cN = 2$ Liouville theory \cite{Hori:2001ax}. 

To study the crosscap/boundary states from the modular bootstrap approach, we need to know the modular transformation of the open string characters. In this approach, we consider the $\cN = 2$ Liouville theory as the CFT which has only $\cN = 2$ super Virasoro symmetry, and we decompose the modular transformed characters in the closed string channel accordingly. The general Liouville theory is a noncompact CFT and possibly has a hidden symmetry other than (super)Virasoro algebra, which contributes to the solvability of the theory (the duality invariance {\it etc}). Nevertheless, it has been believed that the usage of the (super)Virasoro symmetry is enough to solve the crosscap/boundary states problem from the modular bootstrap approach. Actually, the correctness of the modular bootstrap method can be confirmed in some cases by independent conformal bootstrap calculations.

To begin with, let us review the (open string) character of a general $\cN =2$ SCFT. We set $q = e^{2\pi i\tau} $ and $y = e^{2\pi iz}$ as usual. The character of the massive matter representation is 

\begin{equation}
\mathrm{ch}^{(\sNS)}(h,Q;\tau,z) = q^{h-\frac{\hat{c}-1}{8}}y^Q \frac{\theta_3(\tau,z)}{\eta(\tau)^3} \ .
\end{equation}

The character of the massless matter representation is
\begin{equation}
\mathrm{ch}_{M}^{(\sNS)}(Q;\tau,z) = q^{\frac{|Q|}{2}-\frac{\hat{c}-1}{8}}y^{Q} \frac{1}{1+y^{\mathrm{sgn}(Q)}q^{\frac{1}{2}}} \frac{\theta_3(\tau,z)}{\eta(\tau)^3} \ .
\end{equation}

The character of the graviton representation (identity operator) is 
\begin{equation}
\mathrm{ch}_{G}^{(\sNS)}(\tau,z) = q^{-\frac{\hat{c}-1}{8}} \frac{1-q}{(1+yq^{\frac{1}{2}})(1+y^{-1}q^{\frac{1}{2}})} \frac{\theta_3(\tau,z)}{\eta(\tau)^3} \ .
\end{equation}
More general characters for unitary representations are given by the spectral flow of those presented above. In the $\tNS$-sector, we have $\mathrm{ch}_{*}^{(\stNS)}(*;\tau,0)=\mathrm{ch}_{*}^{(\sNS)}(*;\tau,\frac{1}{2})$, where $*$ denotes the abbreviation of corresponding arguments distinguishing their characters. These open string characters have been used to derive the various boundary states for the $\cN =2$ Liouville theory \cite{Eguchi:2003ik,Ahn:2003tt}.

To extend these results to the crosscap states of the $\cN =2$ Liouville theory, we need to use the $\Omega$ inserted characters (M\"obius strip amplitudes). We define the world sheet orientation reversal operator $\Omega$ as
\begin{equation}
\Omega O_r\Omega^{-1} = e^{2\pi i r} O_r \ ,
\end{equation}
where $O_r$ takes either $L_n, G^+_r, G^-_r$ or $J_n$. Because of the phase ambiguity of the half-integral mode, there is another possibility:
\begin{equation}
\widetilde{\Omega} O_r\widetilde{\Omega}^{-1} = e^{-2\pi i r} O_r \ .
\end{equation}
Actually, the both are needed to obtain the GSO projection so that we can assign definite $\Omega$ eigenvalues $\pm 1$ to each physical state.

Then the massive character with the $\Omega$ insertion is given by (we set $z=0$)
\begin{equation}
\mathrm{ch}_\Omega^{(\sNS)}(h,Q;\tau) = q^{h-\frac{\hat{c}-1}{8}} e^{\frac{\pi i}{8}}\frac{\theta_3(\tau+\frac{1}{2})}{\eta(\tau+\frac{1}{2})^3} \ .
\end{equation}
The massless character with the $\Omega$ insertion is
\begin{equation}
\mathrm{ch}_{M,\Omega}^{(\sNS)}(Q;\tau) = q^{\frac{|Q|}{2}-\frac{\hat{c}-1}{8}} \frac{1}{1+iq^{\frac{1}{2}}} e^{\frac{\pi i}{8}}\frac{\theta_3(\tau+\frac{1}{2})}{\eta(\tau+\frac{1}{2})^3} \ .
\end{equation}
The graviton character with the $\Omega$ insertion is 
\begin{equation}
\mathrm{ch}_{G,\Omega}^{(\sNS)}(\tau) = q^{-\frac{\hat{c}-1}{8}} \frac{1+q}{(1+iq^{\frac{1}{2}})^2} e^{\frac{\pi i}{8}}\frac{\theta_3(\tau+\frac{1}{2})}{\eta(\tau+\frac{1}{2})^3} \ .\label{gravom}
\end{equation}
For illustration, we show the lower few levels of graviton character with the $\Omega$ insertion:
\begin{equation}
q^{\frac{\hat{c}}{8}}\mathrm{ch}_{G,\Omega}^{(\sNS)}(\tau) = 1-q-i2q^{\frac{3}{2}}+3q^2 +\cdots \ ,
\end{equation}
The appearance of $i$ indicates that it should be cancelled by combining $\widetilde{\Omega}$ inserted characters (GSO projection) in order to obtain a real partition function.

The massive character with the $\widetilde{\Omega}$ insertion is
\begin{equation}
\mathrm{ch}_{\widetilde{\Omega}}^{(\sNS)}(h,Q;\tau) = q^{h-\frac{\hat{c}-1}{8}} e^{-\frac{\pi i}{8}}\frac{\theta_3(\tau-\frac{1}{2})}{\eta(\tau-\frac{1}{2})^3}\ .
\end{equation}
Note that $\theta_3(\tau-\frac{1}{2}) = \theta_4(\tau+\frac{1}{2})$ and  $\eta(\tau-\frac{1}{2}) = e^{-\frac{\pi i}{12}}\eta(\tau+\frac{1}{2})$. The massless character with the $\widetilde{\Omega}$ insertion is
\begin{equation}
\mathrm{ch}_{M,\widetilde{\Omega}}^{(\sNS)}(Q;\tau) = q^{\frac{|Q|}{2}-\frac{\hat{c}-1}{8}} \frac{1}{1-iq^{\frac{1}{2}}} e^{-\frac{\pi i}{8}}\frac{\theta_3(\tau-\frac{1}{2})}{\eta(\tau-\frac{1}{2})^3} \ .
\end{equation}
Finally, the graviton character with the $\widetilde{\Omega}$ insertion is
\begin{equation}
\mathrm{ch}_{G,\widetilde{\Omega}}^{(\sNS)}(\tau) = q^{-\frac{\hat{c}-1}{8}} \frac{1+q}{(1-iq^{\frac{1}{2}})^2} e^{-\frac{\pi i}{8}}\frac{\theta_3(\tau-\frac{1}{2})}{\eta(\tau-\frac{1}{2})^3} \ .\label{gravtom}
\end{equation}
For the $\tNS$-sector, it is useful to notice $\mathrm{ch}_{*,\widetilde{\Omega}}^{(\sNS)}(*;\tau) = \mathrm{ch}_{*,\Omega}^{(\stNS)}(*;\tau) $.

Although we are interested in the continuum limit, it is useful to define the extended massless character with the $\Omega$ insertion\footnote{One may notice that this definition does not make sense because if $N(m+\frac{r}{N})$ is odd, this character does not have a sensible $\Omega$ projection. In the continuum limit, however, this contribution does not survive. See also the next footnote.}
\begin{equation}
\chi_{M,\Omega}^{(\sNS)}(r,s;\tau) = \sum_{m\in \bsz} \frac{q^{(s-K)(m+\frac{2r+1}{2N})}}{1+iq^{N(m+\frac{2r+1}{2N})}}q^{NK(m+\frac{2r+1}{2N})^2}e^{\frac{\pi i}{8}}\frac{\theta_3(\tau+\frac{1}{2})}{\eta(\tau+\frac{1}{2})^3} \ ,\label{masc}
\end{equation}
where $ \hat{c} = 1+ \frac{2K}{N}$.
Continuum limit ($N \to \infty$, $K \to \infty$) will be taken while keeping
\begin{equation}
r = n \ , \ \ \ \frac{j}{N} = \omega \ , \ \ \ \frac{s}{N} = \lambda \ , \ \ \ \frac{2K}{N} = \cQ^2 \ .
\end{equation}

We also define the spectral-flow summed massless character in the continuum limit as \footnote{A difficulty here is that we do not have the sensible $\Omega$ projection if $r'$ is an odd integer. One way to avoid this problem is to use $\frac{q^{(r'+\frac{1}{2})\omega}}{1+e^{\pi i(r'+\frac{1}{2})}q^{r'+\frac{1}{2}}}$ instead, but this definition is inconvenient for the modular transformation.}
\begin{equation}
\mathrm{Ch}_{M,\Omega}^{(\sNS)}(\omega;\tau) = \sum_{r'\in \bsz}\frac{q^{(r'+\frac{1}{2})\omega}}{1+iq^{r'+\frac{1}{2}}} q^{\frac{\cQ^2(2r'+1)^2}{8}}e^{\frac{\pi i}{8}}\frac{\theta_3(\tau+\frac{1}{2})}{\eta(\tau+\frac{1}{2})^3} \ , \label{quattro}
\end{equation}
where (unspectral-flowed) $U(1)$ charge is related to $\omega$ via $Q=\omega + \frac{\cQ^2}{2}$. This definition will become convenient later.

The modular transformation of the $\Omega$ inserted character or the M\"obius strip amplitude is given by $\tau' = -\frac{1}{4\tau}$ ($q' = e^{2\pi i\tau'}$), which amounts to $TSTTS$ (see Appendix \ref{sec:A}). To utilize the modular bootstrap method to derive the crosscap states, it is necessary to know the transformation property of the graviton representation which contains only the identity representation. We will use the notation $\mathrm{ch}^{(\sNS)}(p,\omega;iT) = \mathrm{ch}^{(\sNS)}(h = \frac{p^2}{2}+\frac{\omega^2}{2\cQ^2}+\frac{\cQ^2}{8},Q=\omega;iT)$.

\begin{align}
\mathrm{ch}_{G,\Omega}^{(\sNS)}(-\frac{1}{4\tau}) &= \frac{i}{2\cQ}\int_0^\infty dp'\int_{-\infty}^\infty d\omega'  \left[\left\{ e^{-\frac{\pi \cQ p'}{2}} \frac{1}{1+i e^{{\pi}(\frac{p'}{\cQ}+i\frac{\omega'}{\cQ^2})}} +e^{\frac{\pi \cQ p'}{2}} \frac{1}{1+i e^{{\pi}(-\frac{p'}{\cQ}-i\frac{\omega'}{\cQ^2})}} \right.\right. \cr
&+\left.e^{-\frac{\pi \cQ p'}{2}} \frac{1}{1+i e^{{\pi}(\frac{p'}{\cQ}-i\frac{\omega'}{\cQ^2})}} + e^{\frac{\pi \cQ p'}{2}} \frac{1}{1+i e^{{\pi}(-\frac{p'}{\cQ}+i\frac{\omega'}{\cQ^2})}} -e^{-\frac{\pi \cQ p'}{2}}-e^{\frac{\pi \cQ p'}{2}}\right\} \cr
 &\times \left. \mathrm{ch}_{\widetilde{\Omega}}^{(\sNS)}(p',\omega';\tau) \right] + \text{discrete terms}\ .\label{moduo}
\end{align}
We will show the contribution from the discrete terms (contribution from the spectral-flow summed massless characters) in Appendix \ref{sec:B}. Here, we just say that the M\"obius strip amplitude for the discrete terms is proportional to $e^{i\pi \frac{\omega'}{2}}$. The argument of the modular transformed function for the discrete terms is the {\it half} of that in the case of the cylinder, which is typical of the M\"obius strip amplitude.

 We will defer the complete derivation to Appendix \ref{sec:B}, but the heuristic derivation of the modular transformation for the continuum part (contribution from the massive representation) \eqref{moduo} goes as follows. First, we rewrite the $\Omega$ inserted character for the graviton representation as
\begin{align}
\mathrm{ch}_{G,\Omega}^{(\sNS)}(-\frac{1}{4\tau}) &= {q'}^{-\frac{\cQ^2}{8}}\left[\sum_{n=0}^\infty (-i)^n 2 {q'}^{\frac{n}{2}}-1 \right]e^{\frac{\pi i}{8}}\frac{\theta_3(-\frac{1}{4\tau} + \frac{1}{2})}{\eta(-\frac{1}{4\tau} + \frac{1}{2})^3} \cr
& = \left[\sum_{n=0}^\infty (-i)^n 2 q'^{\frac{p_n^2}{2}+\frac{\omega_n^2}{2\cQ^2}}-{q'}^{-\frac{\cQ^2}{8}} \right]e^{\frac{\pi i}{8}}\frac{\theta_3(-\frac{1}{4\tau} + \frac{1}{2})}{\eta(-\frac{1}{4\tau} + \frac{1}{2})^3} \ , 
\end{align}
where $p_n = i(\frac{n}{\cQ}-\frac{\cQ}{2}) $ and $\omega_n = n$. By using the modular transformation for the massive character with the $\Omega$ insertion \eqref{modum} and taking the formal resummation, we can obtain the formal modular transformation for the graviton representation.\footnote{However, this method misses the contribution from the massless character, which comes from the existence of the poles when we change the contour of the integration \cite{Miki:1989ri,Eguchi:2003ik}. The complete calculation will be presented in Appendix \ref{sec:B}.}.

\begin{align}
\mathrm{ch}_{G,\Omega}^{(\sNS)}(-\frac{1}{4\tau})  &= \frac{i}{\cQ}\int_0^\infty dp' \int_{-\infty}^\infty d\omega' \left[\sum_{n=0}^\infty (-i)^n2\cosh[\pi p'(\frac{n}{\cQ}-\frac{\cQ}{2})]\cos[\pi\frac{\omega'n}{\cQ^2}] \right. \cr 
&-\left. \cosh[-\frac{\pi p'\cQ}{2}] \right] \mathrm{ch}_{\widetilde \Omega}^{(\sNS)}(p',\omega';\tau) \cr
&= \text{continuum part of (\ref{moduo})} \ .
\end{align}

\sectiono{Crosscap States in $\mathcal{N}=2$ Liouville Theory}\label{sec:3}
In this section, we derive the crosscap states in $\mathcal{N}=2$ Liouville Theory from the modular bootstrap method. In subsection \ref{semic}, we compare our results with the semiclassical minisuperspace approximation. We also check the consistency with the modular bootstrap equation in subsection \ref{sec:con}. We study various overlaps with other branes in subsection \ref{sec:33}.

\subsection{Modular bootstrap}\label{sec:31}
Let us introduce the Ishibashi states \cite{Ishibashi:1988kg} for the $\cN =2$ Liouville theory. For the  boundary states, there are two possible choices:
\begin{align}
 & \text{A-type}: \ \ \: (J_n -\bar{J}_{-n})|B,\eta\rangle, \ \ (G_r^{\pm} - \eta i\bar{G}^{\mp}_{-r} )|B ,\eta\rangle = 0 \cr
 & \text{B-type}: \ \ \:(J_n +\bar{J}_{-n})|B,\eta\rangle, \ \ (G_r^{\pm} - \eta i\bar{G}^{\pm}_{-r}) |B ,\eta\rangle = 0 \ ,
\end{align}
together with 
\begin{equation}
(L_n -\bar{L}_{-n})|B,\eta\rangle, \ \ (G_r - \eta i\bar{G}_{-r} )|B, \eta\rangle = 0 \ ,
\end{equation}
where $G = G^+ + G^-$ is the $\cN =1$ supercurrent. $\eta = \pm$ denotes the spin structure of the boundary states and the proper combination of these sectors enforces the GSO projection. In contrast to some peculiarity in the $\cN =1$ Liouville theory, $\eta$ dependence is trivial in the $\cN =2$ Liouville theory because $|B,+\rangle = e^{i\frac{\pi}{2}(J'_0+\bar{J}'_0)} |B,-\rangle$, where $J'$ is the oscillator part of the $U(1)$ charge.\footnote{$J'$ coincides with $J$ if we are dealing with the supersymmetric case due to the $U(1)$ charge integrality condition.} Therefore we only consider $\eta = +$ sector. In addition, we will only consider the NS-sector of the A-type boundary condition in the following (R-sector will be obtained by the spectral flow).

As the continuum part of the Ishibashi states, we use the following notation and normalization
\begin{equation}
_B\langle\langle p,\omega|e^{-\pi TH^{(c)}}|p',\omega'\rangle\rangle_B = \delta(p-p')\cQ\delta(\omega-\omega') \mathrm{ch}^{(\sNS)}(p,\omega;iT) \ ,
\end{equation}
where $H^{(c)} = L_0 + \bar{L}_0 -\frac{c}{12}$ is the closed string Hamiltonian, and $\mathrm{ch}^{(\sNS)}(p,\omega;iT) = \mathrm{ch}^{(\sNS)}(h = \frac{p^2}{2}+\frac{\omega^2}{2\cQ^2}+\frac{\cQ^2}{8},Q=\omega;iT)$. 
The contribution from the discrete part and the range of the parameter of these Ishibashi states (massless matter sector) are well-discussed in the literature \cite{Eguchi:2003ik,Eguchi:2004yi} and we will not delve into them here.

In the same manner, we can construct the crosscap Ishibashi states from the crosscap condition
\begin{align}
 & \text{A-type}: \ \ \: (J_n -(-1)^n\bar{J}_{-n})|C,\eta\rangle, \ \ (G_r^{\pm} - e^{ir\pi}\eta i\bar{G}^{\mp}_{-r}) |C ,\eta\rangle = 0 \cr
 & \text{B-type}: \ \ \:(J_n +(-1)^n\bar{J}_{-n})|C,\eta\rangle, \ \ (G_r^{\pm} - e^{ir\pi}\eta i\bar{G}^{\pm}_{-r} )|C, \eta\rangle = 0 \ ,
\end{align}
together with 
\begin{equation}
(L_n -(-1)^n\bar{L}_{-n})|C,\eta\rangle, \ \ (G_r - e^{ir\pi}\eta i\bar{G}_{-r}) |C ,\eta\rangle = 0 \ .
\end{equation}
An ambiguity of the half integral phase $e^{ir\pi}$ can be absorbed into $\eta$. In the following we concentrate on the A-type crosscap states.
As a basis of the continuum Ishibashi states, we take the following normalization 
\begin{align}
_B\langle\langle p,\omega,\pm|e^{-\pi TH^{(c)}}|p',\omega',\mp\rangle \rangle_C &= \delta(p-p')\cQ\delta(\omega-\omega') \mathrm{ch}_{\Omega}^{(\sNS)}(p,\omega;iT) \cr
_B\langle\langle p,\omega,\pm|e^{-\pi TH^{(c)}}|p',\omega',\pm\rangle \rangle_C &= \delta(p-p')\cQ\delta(\omega-\omega') \mathrm{ch}_{\widetilde{\Omega}}^{(\sNS)}(p,\omega;iT) \cr
_C\langle\langle p,\omega,\pm|e^{-\pi TH^{(c)}}|p',\omega' ,\pm\rangle\rangle_C &= \delta(p-p')\cQ\delta(\omega-\omega') \mathrm{ch}^{(\sNS)}(p,\omega;iT) \cr
_C\langle\langle p,\omega,\pm|e^{-\pi TH^{(c)}}|p',\omega',\mp\rangle \rangle_C &= \delta(p-p')\cQ\delta(\omega-\omega') \mathrm{ch}^{(\stNS)}(p,\omega;iT) \ .
\end{align}
We note that $|C,+\rangle = e^{i\frac{\pi}{2}(J'_0+\bar{J}'_0)} |C,-\rangle$ again.

The basic assumption of the modular bootstrap for the crosscap states (see {\it e.g.} \cite{Zamolodchikov:2001ah,Hikida:2002bt} for the bosonic case) is that the M\"obius strip amplitude for the Class 1 brane contains only the $\Omega$ inserted character of the identity representation. We write the Cardy crosscap states as
\begin{equation}
\langle C-| = \frac{2}{\cQ}\int_0^\infty dp \int_{-\infty}^{\infty} d\omega \Psi_c(p,\omega) _C\langle\langle p,\omega,-| \ . \label{cardy}
\end{equation}
Then the modular bootstrap assumption states that
\begin{equation}
\langle C-|e^{-\pi TH^{(c)}}|B,O,+\rangle = \mathrm{ch}_{G,\Omega}^{(\sNS)}(it) \ , 
\end{equation}
where the right hand side is the open string character with $T\equiv 1/4t$. The Cardy boundary state $|B,O,+\rangle$ for the identity representation (the Class 1 brane) is well-known \cite{Eguchi:2003ik,Ahn:2003tt} and its basic property is
\begin{equation}
\langle B,O,+|e^{-\pi TH^{(c)}}|B,O,+\rangle = \mathrm{ch}^{(\sNS)}_{G}(it) \ .
\end{equation}

In the Ishibashi state basis, we can write it as
\begin{equation}
\langle B,O,+| = \frac{2}{\cQ}\int_0^\infty dp \int_{-\infty}^{\infty} d\omega \Psi_O(p,\omega) _B\langle\langle p,\omega,+| \ .\label{ishiid}
\end{equation}

From the modular bootstrap method, we obtain
\begin{equation}
\Psi_c(p,\omega) = \Psi_O(-p,\omega)^{-1} P(p,\omega)\ ,
\end{equation}
where modular transformation matrix $P(p,\omega)$ is given by \eqref{moduo}, or more precisely
\begin{align}
P(p,\omega) = \frac{i\cosh(\frac{\pi p}{Q})\left[-i\cosh(\frac{\pi \cQ p}{2})\cos(\frac{\pi \omega}{\cQ^2})+\sinh(\frac{\pi p}{\cQ})\sinh(\frac{\pi Qp}{2})\right]}{\cosh(\frac{2\pi p}{\cQ})+\cos(\frac{2\pi\omega}{\cQ^2})} \ ,\label{pmat}
\end{align}
and the Class 1 (identity representation) brane wavefunction is given by\footnote{The reason we use the complex conjugated wavefunction from the convention in \cite{Eguchi:2003ik} lies in our definition \eqref{cardy},\eqref{ishiid}. This is more natural when we compare it with the one-point function on the real projective plane $\mathbf{RP}_2$. See also section \ref{semic}.}
\begin{equation}
\Psi_O(-p,\omega)^{-1} = \cQ\frac{\Gamma(i\cQ p)\Gamma(1+i\frac{2p}{\cQ})}{\Gamma(\frac{1}{2}-\frac{\omega}{\cQ^2}+i\frac{p}{\cQ}) \Gamma(\frac{1}{2}+\frac{\omega}{\cQ^2}+i\frac{p}{\cQ})} \ ,\label{abns}
\end{equation}
where we have set $\mu=1$ for simplicity, and we continue to use this simplification in the following, but the dependence of $\mu$ can be easily restored from the KPZ scaling law. The discrete part can be obtained similarly.

For a consistency check, let us study the reflection property of the one-point function on $\mathbf{RP}_2$, which is equivalent to the crosscap wavefunction. Since $P(p,\omega)$ is an even function of $p$, the reflection property $p \to -p$ only depends on $\Psi_O(-p)^{-1}$. Thus, it is easy to see that the one-point function on the $\mathbf{RP}_2$ satisfies the desired reflection property \cite{Baseilhac:1998eq} as do the Class 2,3 one-point functions on the disk:
\begin{align}
\Psi_c(p,\omega) = R(p,\omega) \Psi_c(-p,\omega)\ ,
\end{align}
where the reflection amplitude is given by
\begin{equation}
R(p,\omega) = \frac{\Gamma(i\cQ p)\Gamma(1+i\frac{2p}{\cQ})\Gamma(\frac{1}{2}+\frac{\omega}{\cQ^2}-i\frac{p}{\cQ})\Gamma(\frac{1}{2}-\frac{\omega}{\cQ^2}-i\frac{p}{\cQ})}{\Gamma(-i\cQ p)\Gamma(1-i\frac{2p}{\cQ})\Gamma(\frac{1}{2}+\frac{\omega}{\cQ^2}+i\frac{p}{\cQ})\Gamma(\frac{1}{2}-\frac{\omega}{\cQ^2}+i\frac{p}{\cQ})} \ . \label{refrec}
\end{equation}

We end this subsection by presenting one technical comment. In the calculation above, we have implicitly assumed the existence of the complex conjugated contribution to the M\"obius strip amplitude. We have derived the crosscap wavefunction $\Psi_c(p,\omega)$ from the dual open channel NS-sector. However, it should be equally possible to derive it from the open channel $\tNS$-sector. In that case, because of the difference in the degenerate characters between \eqref{gravom} and \eqref{gravtom}, the modular transformation matrix $P^{(\stNS)}(p,\omega)$ becomes the complex conjugation of \eqref{pmat}. To be compatible with the NS-sector calculation, we should understand it as 
\begin{equation}
\Psi_c(p,\omega)^* = \Psi_c(-p,\omega) = P^{(\stNS)}(p,\omega) \Psi_O(p,\omega)^{-1} \ ,
\end{equation}
which amounts to
\begin{equation}
\langle B,O,-|e^{-\pi TH^{(c)}}|C,-\rangle = \mathrm{ch}_{G,\Omega}^{(\stNS)}(it) = \mathrm{ch}_{G,\widetilde\Omega}^{(\sNS)}(it) \ .
\end{equation}

\subsection{Semiclassical limit}\label{semic}
The results obtained above can be checked semiclassically by taking the limit $\cQ \to \infty$. In this limit, the continuum part of the crosscap wavefunction behaves as
\begin{equation}
\Psi_c(p,\omega) \sim \Gamma(i\cQ p) \cosh(\frac{\pi \cQ p}{2}) \ ,\label{semi}
\end{equation}
On the other hand, we can use the minisuperspace approximation to study the semiclassical limit. Setting $\omega = 0$ and assuming that fermionic fields are identically zero in this limit, the Sch\"odinger-like wave equation for the zeromode $\varphi_0$\footnote{In the CFT calculation, we omit the potential term for $\varphi_0$ because it only contributes to the contact terms (see {\it e.g.} \cite{Hosomichi:2004ph} for discussions in the context of the $\cN =2$ Liouville theory). An interesting subtlety here is that to reproduce the CFT result from the semiclassical analysis, we should keep this term. Indeed this potential term is the main source of the semiclassical reflection as we will see.} is given by
\begin{equation}
\left[-\frac{1}{2}\frac{\partial^2}{\partial\varphi_0^2}+2\pi\mu_r^2 e^{\frac{2}{\cQ}\varphi_0}\right]\Phi_p(\varphi_0) = \frac{p^2}{2}\Phi_p(\varphi_0)\ .
\end{equation}
The solution is \cite{Fateev:2000ik,Ahn:2002sx}
\begin{equation}
\Phi_p(\varphi_0) = \frac{(\pi\mu_r^2 \cQ^2)^{-\frac{i\cQ p}{2}}}{\Gamma(-i\cQ p)} K_{i \cQ p} (2\sqrt{\pi\mu_r^2 \cQ^2}e^{\frac{\varphi_0}{\cQ}}) \ ,
\end{equation}
which has the correct reflection property as has been discussed in \cite{Ahn:2002sx}.
To obtain the semiclassical crosscap wavefunction, we calculate the overlap between the semiclassical crosscap state.\footnote{If we calculate the overlap with the semiclassical Class 2 boundary state $\sim e^{-\mu_b^2 e^{\varphi_0/\cQ}}$ we reproduce the semiclassical boundary wavefunction \cite{Ahn:2003tt}.} 
\begin{equation}
\Psi_c(p) \sim \int d\varphi_0 \Phi_c(\varphi_0) \Phi_p(\varphi_0) \ .
\end{equation}
In the minisuperspace approximation, the crosscap condition reduces to imposing the zero momentum condition on $\Phi_c(\varphi_0)$ (see \cite{Gomis:2003vi} for the discussion on the bosonic Liouville theory). Hence $\Phi_c(\varphi_0) = const$ in this limit. Performing the integration, we have
\begin{equation}
\Psi_c(p) \sim \Gamma(i \cQ p) \cosh(\frac{\pi \cQ p}{2}) \ ,
\end{equation}
which exactly equals to what we have derived from the $\cQ \to \infty$ limit of the modular bootstrap results \eqref{semi}.

\subsection{Conformal bootstrap}\label{sec:con}
Another consistency check of the proposed crosscap states is the conformal bootstrap method. In this subsection, we obtain a functional equation which should be satisfied by the one-point functions on the projective plane from the conformal bootstrap equation and show that the crosscap states obtained in section \ref{sec:31} indeed satisfy it. Similar analysis was done for the bosonic unoriented Liouville theory in \cite{Hikida:2002bt}.

Our starting point is the following auxiliary two-point function on the projective plane.
\begin{eqnarray}
\langle N_{\alpha}(z_1,\bar{z}_1)R^+_{-\cQ/4}(z_2,\bar{z}_2)\rangle_{\mathbf{RP}_2} = \frac{(1+z_2\bar{z}_2)^{2\Delta_{\alpha}-2\Delta_{-\cQ/4}}}{|1+z_1\bar{z}_2|^{4\Delta_{\alpha}}} G_\alpha(\eta) \ ,
\end{eqnarray}
where  $N_{\alpha}$ denotes the charge neutral operator in the NS sector with $ip = 2\alpha -\frac{\cQ}{2}$, $\omega = 0$, and $R^{+}_{-\cQ/4}$ denotes the simplest degenerate operator in the (spin $+$) R sector (see {\it e.g.} \cite{Ahn:2002sx,Ahn:2003tt,Ahn:2004qb}). $\Delta_{\alpha}$ is the corresponding  conformal weight and $\eta$ is the crossratio:
\begin{equation}
\eta \equiv \frac{|z_1-z_2|^2}{(1+z_1\bar{z}_1)(1+z_2\bar{z}_2)} \ .
\end{equation}

The basic observation of the conformal bootstrap is that we can write down this two-point function in two different ways by the conformal channel duality. We first introduce the one-point function on the projective plane
\begin{equation}
\langle R^+_{\alpha}(z,\bar{z})\rangle_{\mathbf{RP}_2} = \frac{U^{\sR}(\alpha)}{|1+z\bar z|} \ .
\end{equation}
Then the direct OPE $z_1 \to z_2$ (or $\eta \to 0$) yields the following relation\footnote{The OPE contains only two terms because of the degenerate operator $R^+_{-\cQ/4}$.}
\begin{equation}
G_\alpha(\eta)  = U^{\sR} (\alpha-\frac{\cQ}{4}) F_+(\eta) + C U^{\sR}(\alpha + \frac{\cQ}{4}) F_-(\eta) \ ,\label{direct}
\end{equation}
where the conformal blocks can be written as
\begin{align} 
F_+(\eta) &= \eta^{\cQ\alpha}(1-\eta)^{\cQ\alpha} F(2\cQ\alpha,1-\cQ^2+2\cQ\alpha,1-\frac{\cQ^2}{2}+2\cQ \alpha;\eta) \cr
F_-(\eta) &= \eta^{\frac{\cQ^2}{2}-\cQ\alpha}(1-\eta)^{\cQ\alpha} F(1-\frac{\cQ^2}{2},\frac{\cQ^2}{2},1+\frac{\cQ^2}{2}-2\cQ \alpha;\eta) \ . 
\end{align}
The structure constant $C$ can be calculated perturbatively from the insertion of the dual (nonchiral) screening operator \eqref{dual} (see {\it e.g.} \cite{Ahn:2002sx,Ahn:2003tt}) as
\begin{equation}
C = \tilde{\mu} \pi\gamma(1+\frac{\cQ^2}{2}) \frac{\Gamma(1-2\cQ\alpha)\Gamma(-\frac{\cQ^2}{2}+2\cQ\alpha)}{\Gamma(1+\frac{\cQ^2}{2}-2\cQ\alpha)\Gamma(2\cQ\alpha)} \ .
\end{equation}

On the other hand, we can also evaluate $G_\alpha(\eta)$ in the cross channel. To do this, we first note that the method of image yields 
\begin{equation}
\langle N_{\alpha}(z_1,\bar{z}_1)R^+_{-\cQ/4}(z_2,\bar{z}_2)\rangle_{\mathbf{RP}_2} \sim \langle N_{\alpha}(z_1)R^+_{-\cQ/4}(z_2)N_{\alpha}(-1/\bar{z}_1)\bar{z}_1^{-2\Delta_{\alpha}}R^+_{-\cQ/4}(-1/\bar{z}_2) \bar{z}_2^{-2\Delta_{-\cQ/4}}\rangle_{S^2} \ .
\end{equation}
Thus the dual channel can be defined in the limit $z_1 \to -1/\bar{z}_2$ (or $\eta \to 1$). This gives another expansion for $G_\alpha(\eta)$:
\begin{equation}
G_{\alpha}(\eta)  = U^{\sR} (\alpha-\frac{\cQ}{4}) F_+(1-\eta) + C U^{\sR}(\alpha + \frac{\cQ}{4}) F_-(1-\eta) \ .\label{crossc}
\end{equation}

Now we can equate \eqref{direct} and \eqref{crossc} by using the inversion formula \eqref{eq:hyperg}. This leads to the functional equation for $U^{\sR} (\alpha)$ known as the conformal bootstrap equation: 
\begin{align}
U^{\sR}(\alpha + \frac{\cQ}{4}) &= \frac{\Gamma(-\frac{\cQ^2}{2}+2\cQ\alpha)\Gamma(1+\frac{\cQ^2}{2}-2\cQ\alpha)}{\Gamma(\frac{\cQ^2}{2})\Gamma(1-\frac{\cQ^2}{2})} U^{\sR}(\alpha + \frac{\cQ}{4})  \cr
&+ \frac{1}{\tilde{\mu} \pi \gamma(1+\frac{\cQ^2}{2})} \frac{\Gamma(1+\frac{\cQ^2}{2}-2\cQ\alpha)\Gamma(1-\frac{\cQ^2}{2}+2\cQ\alpha)}{\Gamma(1-2\cQ\alpha)\Gamma(1-\cQ^2+2\cQ\alpha)}U^{\sR}(\alpha-\frac{\cQ}{4}) \ . \label{confbs}
\end{align}

Let us check whether the crosscap states derived in section \ref{sec:31} by using the modular bootstrap method satisfy this functional equation for the one-point function. As usual in the Liouville theory, we define the $\mathbf{RP}_2$ one-point function for the imaginary momentum operator by the analytic continuation of the crosscap wavefunction with the real momentum:
\begin{equation}
U^{\sR}(\alpha) = \Psi^{(\sR)}_c(ip = 2\alpha -\frac{\cQ}{2}, \omega = 0) \ .
\end{equation}
 By performing the spectral flow, we can read the R sector one-point function from \eqref{pmat} and \eqref{abns}
\begin{equation}
U^{\sR}(\alpha)= \frac{i\cQ}{2} \frac{\cos(2\pi\alpha/\cQ)}{\sin(2\pi \alpha /\cQ)} \sin\left(\frac{4\cQ\alpha-\cQ^2}{4}\right) \frac{\Gamma(4\alpha/\cQ)\Gamma(-\frac{\cQ^2}{2}+2\cQ\alpha)}{\Gamma(-\frac{1}{2}+\frac{2\alpha}{\cQ})\Gamma(\frac{1}{2}+\frac{2\alpha}{\cQ})} \ ,
\end{equation}
which results in
\begin{equation}
\frac{U^{\sR}(\alpha+\frac{\cQ}{4})}{U^{\sR}(\alpha-\frac{\cQ}{4})} = 2(-\cQ^2+ 4 \cQ\alpha)\frac{\Gamma(2\cQ\alpha)\sin(\cQ\alpha\pi)}{\Gamma(-\cQ^2 + 1 + 2\cQ\alpha)\sin(\frac{2\cQ\alpha-\cQ^2}{2}\pi)} \ .
\end{equation}
We can see that this is equivalent to the functional equation \eqref{confbs} by suitably choosing the dual cosmological constant.\footnote{Recall that we have neglected the dependence on the cosmological constant from the KPZ scaling argument.} In this way, we have shown that the crosscap states constructed from the modular bootstrap method are consistent with the conformal bootstrap equation for the neutral sector. Due to the $\cN=2$ world sheet supersymmetry, neutral NS one-point functions also satisfy the conformal bootstrap equation.

\subsection{Other overlaps and Klein bottle}\label{sec:33}
It would be interesting to study the overlap between the crosscap state and other boundary states constructed in the literature \cite{Eguchi:2003ik,Ahn:2003tt}.
Let us first list the boundary wavefunctions of these branes. In the following subsections, we will evaluate overlaps between the crosscap states with these boundary states.

Class 1 (continuum part):
\begin{equation}
\Psi_n(p,\omega) = \frac{1}{\cQ} e^{2\pi i n\omega}\frac{\Gamma(\frac{1}{2}+\frac{\omega}{\cQ^2}-i\frac{p}{\cQ})\Gamma(\frac{1}{2}-\frac{\omega}{\cQ^2}-i\frac{p}{\cQ})}{\Gamma(-i\cQ p)\Gamma(1-i\frac{2p}{\cQ})}  \ .
\end{equation}

Class 2:
\begin{equation}
\Psi_{p',\omega'}(p,\omega) = \cQ e^{2\pi i\frac{\omega\omega'}{\cQ^2}} \cos(2\pi p p') \frac{\Gamma(i\cQ p)\Gamma(1+i\frac{2p}{\cQ})}{\Gamma(\frac{1}{2}+\frac{\omega}{\cQ^2}+i\frac{p}{\cQ})\Gamma(\frac{1}{2}-\frac{\omega}{\cQ^2}+i\frac{p}{\cQ})}  \ .
\end{equation}

Class 3 (continuum part):
\begin{align}
\Psi_{\lambda,n}(p,\omega) = \frac{\cQ}{2} e^{2\pi i\omega\frac{\lambda + \cQ^2 n}{\cQ^2}} \frac{\cosh(2\pi(1+\frac{\cQ^2}{2}-\lambda)\frac{p}{\cQ})+e^{-2\pi i\frac{\omega}{\cQ^2}} \cosh(2\pi(\lambda-\frac{\cQ^2}{2})\frac{p}{\cQ})}{2|\cosh\pi(\frac{p}{\cQ}+i\frac{\omega}{\cQ^2})|^2} \cr 
\times \frac{\Gamma(i\cQ p)\Gamma(1+i\frac{2p}{\cQ})}{\Gamma(\frac{1}{2}+\frac{\omega}{\cQ^2}+i\frac{p}{\cQ})\Gamma(\frac{1}{2}-\frac{\omega}{\cQ^2}+i\frac{p}{\cQ})}  \ .
\end{align}

If we allow non-unitary representations for the open channel spectrum, we have what is called ZZ branes in addition \cite{Ahn:2004qb}.\footnote{ZZ Class 2 brane is essentially the same as the Class 3 brane above.}

ZZ Class 1:
\begin{equation}
\Psi_{m,n,\omega'}(p,\omega) = 2\cQ e^{-2\pi i \frac{\omega\omega'}{\cQ^2}} \sinh(\pi m p\cQ)\sinh(2\pi n p/\cQ) \frac{\Gamma(i\cQ p)\Gamma(1+i\frac{2p}{\cQ})}{\Gamma(\frac{1}{2}+\frac{\omega}{\cQ^2}+i\frac{p}{\cQ})\Gamma(\frac{1}{2}-\frac{\omega}{\cQ^2}+i\frac{p}{\cQ})} \ .
\end{equation}

ZZ Class 3 (continuum part):

\begin{equation}
\Psi_m(p,\omega) = \frac{1}{\cQ} \frac{\sinh(\pi m p\cQ)}{\sinh(\pi p \cQ)} \frac{\Gamma(\frac{1}{2}+\frac{\omega}{\cQ^2}-i\frac{p}{\cQ})\Gamma(\frac{1}{2}-\frac{\omega}{\cQ^2}-i\frac{p}{\cQ})}{\Gamma(-i\cQ p)\Gamma(1-i\frac{2p}{\cQ})}  \ .
\end{equation}

The aim of this subsection to study the overlaps between the crosscap states and these branes. For an orientifold plane to be well-defined, we expect that the overlaps with other branes are schematically given by
\begin{equation}
\langle C,-|e^{-\pi TH^{(c)}}|B,+\rangle = \sum_k \chi^{(\sNS)}_{k,\Omega}(it)\ ,
\end{equation}
where the open string characters with the $\Omega$ insertion in the right hand side should appear also in the self-overlaps:
\begin{equation}
\langle B,+|e^{-\pi TH^{(c)}}|B,+\rangle \supseteq \sum_k \chi^{(\sNS)}_{k}(it)\ .
\end{equation}
In this way, we can ensure the $\Omega$ projected trace in the open string sector as
\begin{equation}
Z_C + Z_M = \mathrm{Tr}_{O} \frac{1+\Omega}{2} e^{-2\pi t H^{(o)}} \ ,\label{consis}
\end{equation}
which means that the open string sector has only $\Omega$ invariant states in the physical spectrum.
\subsubsection{Compact branes}
To begin with, let us study the overlap with Class 1, ZZ Class 1 and ZZ Class 3 branes. These branes have a self-overlap which contains only degenerate characters (hence they are compact branes). Thus, we expect that the overlaps between these branes and the crosscap states constructed in section \ref{sec:31} (M\"obius strip amplitudes) contain the corresponding degenerate characters with the $\Omega$ insertion.

The Class 1 brane has the self-overlap (we denote the brane as $|B,n,\eta\rangle$ and consider only the NS-sector):
\begin{equation}
\langle B, n,+| e^{-\pi T H^{(c)}} |B, n,+\rangle = \mathrm{ch}_{G}^{(\sNS)}(it) \ ,
\end{equation}
which is {\it independent} of $n$. However, the overlap with the Class 1 brane with the crosscap state {\it necessarily} depends on $n$:
\begin{align}
\langle C, -| e^{-\pi T H^{(c)}} | B,n,+\rangle &= \frac{2}{\cQ}\int_0^\infty dp\int_{-\infty}^{\infty} d\omega e^{-2\pi i n\omega} P(p,\omega)\mathrm{ch}_{\widetilde \Omega}^{(\sNS)}(p,\omega;iT) \cr
&= \mathrm{ch}_{G,\Omega}^{(\sNS)}(2n;it) \ ,
\end{align}
where $\mathrm{ch}_{G,2n,\Omega}^{(\sNS)}(it)$ is the $\Omega$ inserted character of the $(2n)$ spectral-flowed graviton representation. Thus, it may seem that we do not have a desired $\Omega$ projection properties \eqref{consis}. One way to avoid this problem is to use the combination of the boundary states $|B, n,+\rangle + |B,-n,+\rangle$ as a basic block. In this case, we have a sensible M\"obius strip amplitude because
\begin{equation}
\langle B, -n,+| e^{-\pi T H^{(c)}} |B, n,+\rangle = \mathrm{ch}_{G}^{(\sNS)}(2n;it) \ .
\end{equation}
It would be an interesting problem to investigate this condition under the physical stringy motivated situation.

Next, we consider the overlap with the ZZ Class 1 brane which we denote as $|B_{\text{ZZ}},m,n,\omega',\eta\rangle $. The self-overlap 
\begin{equation}
Z_{m,n,\omega'} = \langle B_{\text{ZZ}},m,n,\omega',+| e^{-\pi T H^{(c)}} |B_{\text{ZZ}},m,n,\omega',+\rangle
\end{equation}
 is given by
\begin{equation}
Z_{m,n,\omega'} = \sum_{k=1,\text{odd}}^{2m-1}\sum_{l=1,\text{odd}}^{2n-1}\left[\chi_{k,l-1,2\omega'}^{(\sNS)} + \chi_{k,l+1,2\omega'}^{(\sNS)}+\chi_{k,l,2\omega'+1}^{(\sNS)}+\chi_{k,l,2\omega'-1}^{(\sNS)}\right] \ , \label{zz1so}
\end{equation}
where the summation is taken over odd integers $k$ and $l$.\footnote{As is pointed out in \cite{Ahn:2004qb}, the right hand side includes {\it reducible} representations, but we will not go into any detail here, which is not the main scope of this paper.} The character of the Class 1 degenerate representation $\chi_{m,n,\omega'}^{(\sNS)}$ is defined as
\begin{equation}
\chi_{m,n,\omega'}^{(\sNS)}(\tau) = \left[q^{-\frac{1}{2}(\frac{m\cQ}{2}+\frac{n}{\cQ})^2}-q^{-\frac{1}{2}(\frac{m\cQ}{2}-\frac{n}{\cQ})^2}\right]q^{\frac{{\omega'}^2}{2\cQ^2}} \frac{\theta_3(\tau)}{\eta(\tau)^3} \ ,
\end{equation}
which has the modular transformation:
\begin{equation}
\chi_{m,n,\omega'}^{(\sNS)}(-\frac{1}{\tau}) = \frac{4}{\cQ}\int_0^\infty dp \int_{-\infty}^{\infty}d\omega e^{-2\pi i \frac{\omega\omega'}{\cQ^2}} \sinh(\pi m p\cQ)\sinh(2\pi n p/\cQ) \mathrm{ch}_{}^{(\sNS)}(p,\omega;\tau) 
\end{equation}
 Calculating the overlap with the crosscap states, we find that we can reproduce some terms of the $\Omega$ inserted characters corresponding to \eqref{zz1so} (for $\omega' = 0$) in the open string channel. Let us illustrate this in the simplest example: $m=n=1$, $\omega' = 0$.
\begin{align}
\langle C,-| e^{-\pi T H^{(c)}} |B_{\text{ZZ}},1,1,0,+\rangle = \frac{2}{\cQ}\int_0^\infty dp \int_{-\infty}^{\infty}d\omega \cr 
i\cosh(\frac{\pi p}{\cQ})\left[-i\cosh(\frac{\pi p\cQ}{2})\cos(\frac{\pi \omega}{\cQ^2}) + \sinh(\frac{\pi p}{\cQ})\sinh(\frac{\pi \cQ p}{2})\right]\mathrm{ch}_{\widetilde{\Omega}}^{(\sNS)}(p,\omega;iT) \ .
\end{align}
The first factor in the parenthesis yields $\chi_{1,1,1,\Omega}^{(\sNS)}$ and $\chi_{1,1,-1,\Omega}^{(\sNS)}$ and the second factor yields $\chi_{1,2,0,\Omega}^{(\sNS)}.$ To see this, it is important to realize that the degenerate character with the $\Omega$ insertion  has a different modular transformation according to the sign of $(-1)^{mn}$ as
\begin{align}
\chi_{m,n,\omega',\Omega}^{(\sNS)} = \frac{2i}{\cQ}\int_0^\infty dp \int_{-\infty}^{\infty}d\omega e^{-\pi i \frac{\omega\omega'}{\cQ^2}} \sinh(\pi m p\cQ/2)\sinh(\pi n p/\cQ) \mathrm{ch}_{\widetilde{\Omega}}^{(\sNS)}(p,\omega;iT)\ ,
\end{align}
if $ \ (-1)^{mn} = +1$, and 
\begin{equation}
\chi_{m,n,\omega',\Omega}^{(\sNS)} = \frac{2i}{\cQ}\int_0^\infty dp \int_{-\infty}^{\infty}d\omega e^{-\pi i \frac{\omega\omega'}{\cQ^2}} \cosh(\pi m p\cQ/2)\cosh(\pi n p/\cQ) \mathrm{ch}_{\widetilde{\Omega}}^{(\sNS)}(p,\omega;iT) \ ,
\end{equation}
if $\ (-1)^{mn} = -1$. This phenomenon, in addition to the halving of the argument, has been observed also in the bosonic Liouville theory \cite{Hikida:2002bt}. However, we find that in more general situations, these overlaps with the crosscap state do not reproduce all the terms corresponding to \eqref{zz1so}. It is a challenging question whether some combinations of the boundary states may cure the situation as in the Class 1 case demonstrated above. Another possibility is that $\Omega$ acts differently on some open spectrums; for instance, the open strings stretching between a brane and its orientifold mirror do not necessarily have an $\Omega$ projected spectrum.

Similarly, we can analyse the overlap with the ZZ Class 3 brane which we denote as $|B_{\text{ZZ}},m,\eta\rangle $. Then the self-overlap is given by
\begin{equation}
Z_{m,\text{ZZ}} = \langle B_{\text{ZZ}},m,\eta| e^{-\pi T H^{(c)}} |B_{\text{ZZ}},m,\eta\rangle  = \sum_{k=1,\text{odd}}^{2m-1} \chi_k^{(\sNS)} \ , \label{zz3so}
\end{equation}
where $\chi_k^{(\sNS)}$ corresponds to the character of the $k$-th Class 3 degenerate representation:
\begin{equation}
\chi_k^{(\sNS)} = q^{-\frac{k^2\cQ^2}{8}} \frac{1-q^{k}}{(1+q^{\frac{k}{2}})^2}\frac{\theta_3(\tau)}{\eta(\tau)^3} \ .
\end{equation}
 We can calculate the overlap with the crosscap state, and we find that the M\"obius strip amplitude reproduces the $\Omega$ ($\widetilde{\Omega}$) inserted version of \eqref{zz3so}.

For example, let us take the simplest $m=2$ case. The overlap with the crosscap states gives 
\begin{align}
\langle C,-| e^{-\pi T H^{(c)}} |B_{\text{ZZ}},2,-\rangle \cr = \frac{i}{2\cQ}\int_0^\infty dp \int_{-\infty}^{\infty}d\omega (e^{\pi \cQ p} + e^{-\pi \cQ p})
\left[\left\{ e^{-\frac{\pi \cQ p}{2}} \frac{1}{1+i e^{{\pi}(\frac{p}{\cQ}+i\frac{\omega}{\cQ^2})}} +e^{\frac{\pi \cQ p}{2}} \frac{1}{1+i e^{{\pi}(-\frac{p}{\cQ}-i\frac{\omega}{\cQ^2})}} \right.\right. \cr
+\left.e^{-\frac{\pi \cQ p}{2}} \frac{1}{1+i e^{{\pi}(\frac{p}{\cQ}-i\frac{\omega}{\cQ^2})}} + e^{\frac{\pi \cQ p}{2}} \frac{1}{1+i e^{{\pi}(-\frac{p}{\cQ}+i\frac{\omega}{\cQ^2})}} -e^{-\frac{\pi \cQ p}{2}}-e^{\frac{\pi \cQ p}{2}}\right\} \left. \mathrm{ch}_{{\Omega}}^{(\sNS)}(p,\omega;iT) \right] \ . \label{zzaa}
\end{align}
This contains $\chi_{1,\widetilde{\Omega}}^{(\sNS)}$ and $\chi_{3,\widetilde{\Omega}}^{(\sNS)}$ as desired because 
\begin{align}
{\chi}_{1,\widetilde{\Omega}}^{(\sNS)} = \frac{-i}{2\cQ}\int_0^\infty dp \int_{-\infty}^{\infty}d\omega 
\left[\left\{ e^{+\frac{\pi \cQ p}{2}} \frac{1}{1+i e^{{\pi}(\frac{p}{\cQ}+i\frac{\omega}{\cQ^2})}} +e^{-\frac{\pi \cQ p}{2}} \frac{1}{1+i e^{{\pi}(-\frac{p}{\cQ}-i\frac{\omega}{\cQ^2})}} \right.\right. \cr
+\left.e^{+\frac{\pi \cQ p}{2}} \frac{1}{1+i e^{{\pi}(\frac{p}{\cQ}-i\frac{\omega}{\cQ^2})}} + e^{-\frac{\pi \cQ p}{2}} \frac{1}{1+i e^{{\pi}(-\frac{p}{\cQ}+i\frac{\omega}{\cQ^2})}} -e^{-\frac{\pi \cQ p}{2}}-e^{\frac{\pi \cQ p}{2}}\right\} \left. \mathrm{ch}_{{\Omega}}^{(\sNS)}(p,\omega;iT) \right] \ ,
\end{align}
and 
\begin{align}
\chi_{3,\widetilde{\Omega}}^{(\sNS)} = \frac{-i}{2\cQ}\int_0^\infty dp \int_{-\infty}^{\infty}d\omega 
\left[\left\{ e^{-3\frac{\pi \cQ p}{2}} \frac{1}{1+i e^{{\pi}(\frac{p}{\cQ}+i\frac{\omega}{\cQ^2})}} +e^{+3\frac{\pi \cQ p}{2}} \frac{1}{1+i e^{{\pi}(-\frac{p}{\cQ}-i\frac{\omega}{\cQ^2})}} \right.\right. \cr
+\left.e^{-3\frac{\pi \cQ p}{2}} \frac{1}{1+i e^{{\pi}(\frac{p}{\cQ}-i\frac{\omega}{\cQ^2})}} + e^{+3\frac{\pi \cQ p}{2}} \frac{1}{1+i e^{{\pi}(-\frac{p}{\cQ}+i\frac{\omega}{\cQ^2})}} -e^{-3\frac{\pi \cQ p}{2}}-e^{3\frac{\pi \cQ p}{2}}\right\} \left. \mathrm{ch}_{{\Omega}}^{(\sNS)}(p,\omega;iT) \right] \ .
\end{align}
Note the sign difference in the term $e^{\pm k\frac{\pi \cQ p'}{2}}$. This is due to the fact that the correct $\Omega$ inserted amplitude is given by
\begin{equation}
\chi_{k,\Omega}^{(\sNS)} = q^{-\frac{k^2\cQ^2}{8}} \frac{1-e^{i\pi k}q^k}{(1+e^{\frac{i\pi}{2}k} q^{\frac{k}{2}})^2} e^{\frac{\pi i}{8}}\frac{\theta_3(\tau+\frac{1}{2})}{\eta(\tau+\frac{1}{2})^3} \ .
\end{equation}
Similarly we can show that the M\"obius strip amplitude for more general cases reproduces the $\Omega$ inserted version of \eqref{zz3so}.\footnote{The choice between $\Omega$ and $\widetilde{\Omega}$ depends on the sign of $(-1)^k$. This was the reason we chose $|B_{\text{ZZ}},2,-\rangle$ in \eqref{zzaa}. Otherwise the wrongly projected character would appear.}

\subsubsection{Noncompact branes}
Let us move on to the analysis of the overlap between the Class 2 branes and the crosscap states. The Class 3 brane can be dealt with almost in the same fashion. As we will see, this overlap is even subtler than the overlaps discussed so far, whose peculiar structure is also found in the bosonic or $\cN =1$ Liouville theory (see \cite{Hikida:2002bt,Nakayama:2003ep,Nakayama:2004vk}). In this case, the self-overlap is given by the integration over the massive characters in the open string channel, and they are noncompact branes. Thus, the central object is the spectral density of the open string.

The density of states (or the Fourier transform of the absolute square of the boundary wave function) bounded between the Class 2 branes (with parameters $p'$ and $\omega'$) is given by
\begin{equation}
\rho(P,W;p',\omega') = \frac{4}{\cQ}\int_0^\infty dp \int_{-\infty}^\infty d\omega \cos(2\pi pP) e^{2\pi i\frac{\omega W}{\cQ^2}} \frac{\cosh(\frac{2\pi p}{\cQ})+\cos(\frac{2\pi \omega}{\cQ^2})}{\sinh(\pi \cQ p)\sinh(\frac{2\pi p}{\cQ})} \cos(2\pi pp')^2 \ . \label{densb}
\end{equation}
On the other hand, the overlap between the Class 2 brane and the crosscap states leads to
\begin{align}
\rho'(P,W;p',\omega') &= \frac{2}{\cQ}\int_0^\infty dp \int_{-\infty}^\infty d\omega \cos(\pi pP)e^{i\pi\frac{\omega W}{\cQ^2}} \cos(2\pi pp') \cr 
&\times \frac{i\cosh(\frac{\pi p}{\cQ})\left[-i\cosh(\frac{\pi p\cQ}{2})\cos(\frac{\pi \omega}{\cQ^2}) + \sinh(\frac{\pi p}{\cQ})\sinh(\frac{\pi \cQ p}{2})\right]}{\sinh(\pi \cQ p)\sinh(\frac{2\pi p}{\cQ})} \ . \label{densc}
\end{align}
Though the distribution in the $W$ direction has the similar structure, the $P$ dependence is different from each other (no matter what value we take as $p'$ which is related to the boundary cosmological constant). Therefore it seems inconsistent with the $\Omega$ projection condition \eqref{consis}. Note that this pathology was also observed in the bosonic or $\cN=1$ Liouville theory \cite{Nakayama:2003ep,Nakayama:2004vk}. One might try to construct a crosscap state which has a correct overlaps with these Class 2 branes (or FZZT branes in the bosonic case), but then we would necessarily end up with the $p'$ dependent crosscap state. Furthermore, in the bosonic case, such a crosscap state is shown to be inconsistent with the conformal bootstrap constraint \cite{Hikida:2002bt,Nakayama:2004vk}. It would be interesting to investigate this problem in the $\cN =2$ case.

One argument to overcome this pathology is the following.\footnote{This was originally suggested by J.~Gomis and A.~Kapustin to the author in the context of the $\cN =1$ Liouville theory in two dimension.} First note that the expressions \eqref{densb} and \eqref{densc} are actually divergent in the infrared limit $p \to 0$. Because of this divergence, the density of states contains a diverging constant part which should have a cutoff as $\sim \log \mu$. This part is the same both for \eqref{densb} and \eqref{densc}, so we may conclude that the ``bulk" part of the open string spectrum has a sensible $\Omega$ projection. At the same time, in the matrix model calculation (if any), the $P$ dependent part of the density of states can be regarded as a nonuniversal term and simply dropped in the double scaling limit.\footnote{We do not say that the finite part is irrelevant in the physical application. On the contrary, this part can be compared with the boundary two-point function to check its consistency.} We do not claim this should be always true, but otherwise we should face this pathology. See also the discussion on the strict modular invariance which is obtained after dividing the infinite volume \cite{Eguchi:2004yi}.

\subsubsection{Klein bottle}

Finally let us briefly discuss the Klein bottle amplitude. The Klein bottle amplitude is given by
\begin{align}
Z_K^{(\sNS)} = \langle C|e^{-\pi T H^{(c)}}| C \rangle = \frac{2}{\cQ}\int_0^\infty dp \int_{-\infty}^\infty d\omega|\Psi_c(p,\omega)|^2 \mathrm{ch}^{(\sNS)}(p,\omega,iT) \ .
\end{align}
After the modular transformation $\tau \to -\frac{1}{2\tau}$, we obtain the density of states for the $\Omega$ inserted closed string:
\begin{equation}
\rho(P,W) = \frac{2}{\cQ}\int_0^\infty dp \int_{-\infty}^\infty d\omega \cos(2\pi pP)e^{2\pi i\frac{\omega W}{\cQ^2}} |\Psi_c(p,\omega)|^2 \ . \label{densn}
\end{equation}
This can be compared with the closed string density of states which can be obtained independently (see {\it e.g.} \cite{Eguchi:2004yi}). Especially, the finite part will be expected to match with the derivative of the logarithm of bulk reflection amplitudes \eqref{refrec} if the $\Omega$ projection persists in the full closed string spectrum. However, it is not straightforward to see the coincidence here. Since the density of states \eqref{densn} suffers an infrared divergence, it may be possible that the $\Omega$ projection only acts on the divergent bulk spectrum as in the $\Omega$ projection of the noncompact branes (see the argument on the strictly modular invariant partition function in \cite{Eguchi:2004yi}). We also note that the same unsolved problem exists in the bosonic Liouville theory \cite{Hikida:2002bt,Nakayama:2004vk}.

\section{Discrete part and topological Landau-Ginzburg approach}\label{sec:top}

The $\cN = 2$ Liouville theory has $\cN = 2$ world sheet superconformal symmetry from its definition. As a consequence, we can perform a topological twist. Since we can regard the $\cN =2 $ Liouville theory as a Landau-Ginzburg theory whose super potential is given by \eqref{chil} with a {\it linear dilaton} background, it is interesting to see whether the powerful topological Landau-Ginzburg calculational method ({\it e.g.} \cite{Vafa:1990mu}) is applicable here. From this motivation, we will concentrate on the B-model and its B-twist of the $\cN = 2$ Liouville theory with the chiral superpotential in this section.\footnote{This should be dual to the $A$-twist of the $\cN =2$ Liouville theory with the dual nonchiral Kahler potential, or the $SL(2,\mathbf{R})/U(1)$ supercoset model.}

As is well-known, the BRST cohomology of the twisted $\cN=2$ algebra is spanned by the chiral representations of the original algebra. Therefore, the relevant physical states are the chiral primary operators whose operator form is symbolically given by $e^{\alpha S} = e^{\alpha (\phi + i Y)}$ in the NS-sector. With the spectral flow, we obtain corresponding R ground states. 

Concerning the crosscap states (or boundary states), the fundamental topological objects are the period integral, or the coupling to the R ground states.
\begin{align}
\Pi_i^\gamma &= \langle i | B_\gamma \rangle \cr
\Pi_i^{\Omega} &= \langle i| C \rangle \ ,
\end{align}
where $i$ denotes the label of the R ground states. {}From the CFT perspective, they are encoded in the discrete part of the boundary wavefunction \eqref{disp},\eqref{disc}. In the case of the Lagrangian A-brane in the topological Landau-Ginzburg model, there is a standard method to evaluate these period integrals \cite{Hori:2000kt,Hori:2000ck}:
\begin{equation}
\Pi_i^\gamma = \int dX_{\gamma} \frac{1}{g_s} \Phi_i(X) \exp(-W(X)) \ , 
\end{equation}
where $\Phi_i(X)$ is the corresponding chiral primary operator and $W(X)$ is the Landau-Ginzburg superpotential. The Lagrangian cycle $\gamma$ is given by the condition $\mathrm{Im} W(X) = const$, whose end point is prescribed to be an extremum of the superpotential. Here we have included the string coupling $g_s$ so that it measures the BPS tension of the brane. The same reasoning applies to the crosscap state \cite{Brunner:2003zm} and we have
\begin{equation}
\Pi_i^\Omega = \int dX_{\gamma_\Omega} \frac{1}{g_s} \Phi_i(X) \exp(-W(X)) \ , \label{nobackd}
\end{equation}
where $\gamma_\Omega$ is given by the orientifold fixed cycle $W(X) = \bar{W}(\bar{X})$. It is reported that these integrals correctly reproduce the coupling of the  Cardy crosscap/boundary states to the chiral primaries (or R ground states) in the $\cN = 2$ minimal model whose Landau-Ginzburg description is well-known \cite{Hori:2000ck,Brunner:2003zm}.

How can we generalize this result to the nontrivial dilaton background? To this question there is a proposal by \cite{Hori:2002cd}, whose argument goes as follows. The nontrivial dilaton background may be regarded simply as a nontrivial position dependant string coupling constant. Thus, the natural extension of the above prescription \eqref{nobackd} is to replace $\frac{1}{g_s}$ with the position dependent coupling constant $\frac{1}{g_s(X)}$. The linear dilaton in our case leads us to the simple expression $\frac{1}{g_s} = e^{-\frac{\cQ}{2} \mathrm{Re} S}$. As discussed in \cite{Hori:2002cd}, a plausible holomorphic extension is given by $\frac{1}{g_s} \to e^{-\frac{\cQ}{2} S}$. In this way, together with the chiral superpotential $W(S) = \mu e^{\frac{1}{\cQ}S}$ for the $\cN =2$ Liouville theory, the period integrals become
\begin{equation}
\Pi_i^\gamma = \int dS_{\gamma} \Phi_i(S) \exp(-\frac{S\cQ}{2}-\mu e^{\frac{1}{\cQ}S}) \ . \label{peris}
\end{equation}
It is worthwhile mentioning that this procedure has been utilized in \cite{Hori:2002cd} for the noncompact Gepner models including two $\cN=2$ Liouville sectors, which becomes the solvable CFT description of the wrapped NS5 branes, reproducing the Seiberg-Witten theory from the world sheet perspective \cite{Klemm:1996bj,Lerche:2000uy}.

{}From the weak coupling limit analysis, we would take chiral primary operators as $\Phi_i(S) = e^{\alpha_i S}$, where $U(1)$ charge $\alpha_i$ is given by the integral multiples of $\frac{1}{\cQ}$ within the unitarity bound. Determining the contour from the condition $\mathrm{Im} W(S) = const$ (boundary states) or $W(S) = \bar{W}(\bar{S})$ (crosscap states), we would be able to reproduce the results of the modular bootstrap method --- especially the discrete term whose conformal bootstrap approach would be difficult.

However, to fully reproduce CFT results such as \eqref{disp},\eqref{disc}, we have to further specify the normalization of the R ground states. This was just what we needed to compare the results from the Landau-Ginzburg approach with the Cardy boundary/crosscap states for the $\cN=2$ minimal models. The problem of determining the normalization of the R ground states (or NS chiral primaries) is closely related to the problem of finding a prescription to obtain two-point correlation function on the sphere from the topological Landau-Ginzburg theory with a {\it nontrivial dilaton background}. We will leave this important problem for the future study and just make some remarks here.\footnote{Recently, the topological Landau-Ginzburg method \cite{Vafa:1990mu} is extended to the correlation function including boundary operators in \cite{Kapustin:2003ga,Kapustin:2003rc,Brunner:2003dc,Herbst:2004ax,Herbst:2004zm} with the B-type boundary condition. It will become a powerful technique if it is also applicable to the boundary $\cN =2$ Liouville theory or even the Landau-Ginzburg theory with $X^{-1}$ superpotential (see the following comments). }

\noindent
{\bf 1. } The Lagrangian condition $\mathrm{Im} W(S) = const$ is rewritten in the component form as 
\begin{equation}
\mu_r \sin\left(\frac{Y}{\cQ}\right) = e^{-\frac{\phi}{\cQ}}  \ . \label{lag}
\end{equation}
This curve, at first sight, looks similar to the hairpin curve, which is the large $N$ limit of the Class 2 brane \cite{Ribault:2003ss,Lukyanov:2003nj,Nakayama:2004yx}. However, a closer look at the precise form will make us realize some differences. First of all, \eqref{lag} extends to the {\it strong} coupling region, whereas the hairpin brane extends to the {\it week} coupling region. Thus, the more natural interpretation of this Lagrangian brane may be the Class 1 (or 3) brane. This is consistent with the boundary state analysis because the Class 2 brane does not couple to the R ground states (NS primary states) at all. 

\noindent
{\bf 2. } Even if we identify this brane as the Class 1 (or 3) brane which is localized in the strong coupling region, there is another subtlety concerning the shape of the brane given by \eqref{lag}. As is discussed in \cite{Nakayama:2004yx}, the Fourier transform of the boundary wavefunction yields the shape of the brane in the semiclassical limit. However, the direct calculation shows that for any wavefunction of the possible D-brane boundary states, the coordinate dependence of the brane shape is constructed only from the combination $\phi \cQ$ and $ Y \cQ$ in the large $N$ ($\cQ \to 0$) limit. Thus, it does not seem to match with the Lagrangian cycle condition \eqref{lag}, which only depends on the combination $\phi/\cQ$ and $ Y/\cQ$. Nevertheless, it is important to note that this large $N$ limit ($\cQ \to 0$) is actually not the semicalssical limit of the chiral Liouville superpotential, rather that of the dual Kahler potential (compare it with the semiclassical limit taken in section \ref{semi}), so there is no logical necessity that the two different limits, hence the brane shapes, coincide. These observations are all consistent, but the further study on this point will clarify the whole structure.

\noindent
{\bf 3. } It is worthwhile mentioning that we can rewrite the period integrals \eqref{peris} as period integrals of the noncompact Landau-Ginzburg model {\it in the trivial dilaton background}. If we make the change of variables as $ X = e^{-\frac{S\cQ}{2}}$, we observe that the measure factor cancels with the linear dilaton term:
\begin{equation}
\Pi_i^\gamma = \int dX_{\gamma'} \Phi_i'(X) \exp(-\mu X^{-\frac{2}{\cQ^2}}) \ , 
\end{equation}
which is nothing but the period integral of the Landau-Ginzburg theory with the superpotential $\mu X^{-\frac{2}{\cQ^2}}$. Thus, the period integral of the $\cN=2$ Liouville theory could be evaluated also from the Landau-Ginzburg theory with the $\mu X^{-\frac{2}{\cQ^2}}$ superpotential without any nontrivial dilaton contribution. Note that the Landau-Ginzburg theory with the negative power superpotential $W = X^{-k}$ is thought to be equivalent to the $SL(2,\mathbf{R})_{k+2}/U(1)$ supercoset model whose central charge is $\hat{c} = 1 + \frac{2}{k}$ \cite{Ooguri:1995wj,Mukhi:1993zb} from the analytic continuation of the well-established fact that the $SU(2)_{k+2}/U(1)$ supercoset model is equivalent to the Landau-Ginzburg model with $W = X^{k}$. However, the $SL(2,\mathbf{R})_{k+2}/U(1)$ model is the mirror dual of the original $\cN =2$ Liouville theory because the central charge $\hat{c} = 1 + \frac{2}{k} = 1 + \cQ^2$ does not change in this process. Therefore, we find that we can extend this chain of dualities to the (topological) open string sector in the weak sense described in \cite{Hori:2000kt}.

In the special case of $\cQ = \sqrt{2}$ (hence $\hat{c} = 1+\sqrt{2}^2 = 3$), which is related to the conifold background \cite{Mukhi:1993zb,Ghoshal:1995wm,Ooguri:1995wj}, interesting structure emerges. In this case, the period integral is given by
\begin{equation}
\Pi_i^\gamma = \int dX_{\gamma'} \Phi_i'(X) \exp(-\mu X^{-1}) \  
\end{equation}
after introducing $ X = e^{-\frac{S}{\sqrt{2}}}$. This extends the result of \cite{Ghoshal:1993qt,Ghoshal:1995wm} --- the conifold, $\hat{c}=3$ $\cN =2$ Liouville theory and the Landau Ginzburg theory with $\mu X^{-1}$ superpotential are all equivalent, to the open string sector. Also note that $\Phi_i'(X)$ is the monomial of $X$ if we take $\Phi(S) = e^{k\frac{S}{\sqrt{2}}}$ with $k \in \bz$ as the Liouville chiral primary operators.

\sectiono{Discussion and Summary}\label{sec:5}
\subsection{Possible applications}\label{appl}
In this subsection, we present possible applications of our result.

\subsubsection{Crosscap states in two dimension and tadpole cancellation}
One of the obvious applications of our results is to construct a fermionic (type 0 or 2) unoriented string theory in two dimension. By setting $\hat{c} =5$ {\it i.e.} $\cQ = 2$, we obtain such a background. Taking account of the duality to the $SL(2,\mathbf{R})/U(1)$ supercoset model, one can regard this background as a two-dimensional fermionic black hole.\footnote{Strictly speaking, this is not true in our setup because we are dealing with the noncompact $U(1)$ (or $Y$) direction. Notice, however, if we Wick rotate it to the Lorentzian black hole, $U(1)$ direction becomes noncompact.} Since there is no other CFT sectors, this background is the simplest setup to use $\cN = 2$ (unoriented) Liouville theory as a part of the string theory.

In this background, we can see that the Class 2,3 branes and the orientifold plane has a diverging tadpole coupling in the NS-sector (take $p \to 0$, $\omega \to 0$ in the boundary/crosscap wavefunction). These can be cancelled by combining these branes and orientifold planes with the suitable Chan-Paton gauge indices. Alternatively saying, the one-loop partition function
\begin{equation}
Z^{(\sNS)} = Z^{(\sNS)}_{C}+Z^{(\sNS)}_{M}+Z^{(\sNS)}_{K} \label{loop}
\end{equation}
may become finite. In fact, the NS tadpole cancels if we combine one Class 2 brane and the orientifold plane. Note that in contrast to the free theory, $\eqref{loop}$ cannot completely vanish due to the subtlety in the $\Omega$ projection on the discrete part (see section 3). This was also observed in the bosonic and $\cN=1$ Liouville theory \cite{Nakayama:2003ep,Nakayama:2004vk}.

In the R-sector, things become more complicated. R-wavefunctions for the boundary/crosscap states are obtained by replacing $\omega_{NS} = \omega_R -\frac{\cQ^2}{2}$. Then we find that the tadpole is less divergent than in the NS-sector --- more precisely constant in the $p,\omega \to 0$ limit both for the noncompact Class 2,3 branes and the orientifold plane. In the case of the brane, this is explained by the fact that the Class 2 brane can be regarded as a brane-antibrane pair asymptotically as has been discussed in \cite{Nakayama:2004yx}.
On the other hand, the one-loop amplitudes are divergent, and the orientifold plane is necessary to cancel this divergence. 

This difference between the NS part and R part which contrasts with the naive free field guess may have a physical explanation. Taking the Fourier transform of the semiclassical wavefunction, we find that the shape of the Class 2 brane is given by the hairpin curve \cite{Ribault:2003ss,Lukyanov:2003nj,Nakayama:2004yx}. In \cite{Nakayama:2004yx}, we have discussed that the Wick-rotated rolling brane has an IR divergence only in the NS-NS sector because it roles into the strong coupling region and the gravitational coupling to branes vanishes there. On the other hand in our hairpin case, the brane extends in the weak coupling region and hence it is natural to suppose that the gravitational coupling to the brane/orientifold dominates over the R-R coupling.

However, we should note that the notion of tadpole or physical state is very formal in the Euclidean two-dimensional setup considered here. One may overcome this difficulty by applying the Wick-rotation of the $Y$ direction \cite{Nakayama:2004yx} or the analytic continuation of the $\phi$ direction with the analytic continuation of the background charge $\cQ$ \cite{Hikida:2004mp}. This procedure is very nontrivial and possibly changes the structure of the boundary/crosscap states itself.

After the introduction of the orientifold plane (with or without D-branes), we can suggest  the matrix model dual for this two-dimensional unoriented string theory. Obvious candidates are the $SO/Sp$ group version of the fermionic KKK matrix model \cite{Kazakov:2000pm,Giveon:2003wn} (for type 0; see also \cite{Gomis:2003vi,Bergman:2003yp,Carlisle:2004jn} in the $\cN =1$ case), or the $SO/Sp$ group version of the supermatrix models \cite{McGreevy:2003dn,Verlinde:2004gt} (for type 2). Further introduction of Class 2,3 branes may add vector degrees of freedom into these setups \cite{Klebanov:2003km,Bergman:2003yp}.

\subsubsection{Orientifold plane in noncompact superstring background}
Another application may be the orientifold plane in the noncompact superstring background. By combining $\cN = 2$ Liouville theory with other compact CFTs such as $\cN=2$ minimal models, we can produce a large class of $\alpha'$ exact string background which preserves the space-time SUSY. If we try to preserve space-time SUSY after introducing orientifold plane (with or without other branes), we need to ensure that the $U(1)$ charge of the open/closed string sector should have integral values. Thus, this setup is beyond the scope of this paper, where we have restricted ourselves to the continuum $U(1)$ charge. Nevertheless, we try to present an outline of the arguments, pointing out necessary modifications.

To apply our results to the supersymmetric orientifold, we should compactify $Y$ direction and sum over the partial spectral flows as has been discussed in \cite{Eguchi:2003ik}. This is nothing but the extended character (with the $\Omega$ insertion) whose modular transformation properties we have reported in Appendix \ref{sec:B}.\footnote{We admit that the extended character with the $\Omega$ insertion we have used in this paper involves  wrongly projected states which go away in the continuum limit. This part should be modified if we try to tackle the problem of supersymmetric orientifolds.} Then we would combine the Ishibashi states with those of the $\cN =2$ minimal models to construct the full Ishibashi states for the crosscap states while taking the GSO projection which guarantees the integral $U(1)$ charge. The supersymmetric crosscap states in the minimal models have been discussed recently in the context of the Gepner models (see {\it e.g.} \cite{Govindarajan:2003vv,Govindarajan:2003vp,Blumenhagen:2003su,Aldazabal:2003ub,Brunner:2004zd}). Inclusion of the noncompact $\cN =2$ Liouville sector extends the realm of the Gepner models into the noncompact Gepner models, where the gravity decouples. We also expect the connection to the local mirror symmetry of the Calabi-Yau orientifold.

Once we construct these crosscap states, we can probe the geometry by looking at the $\widetilde{R}$-sector of the one loop amplitudes. As we have briefly discussed in the last section, the discrete part of the crosscap wavefunction measures the period integral of the orientifold plane. We could also calculate the intersection number of the orientifold planes by calculating the Witten index.

The orientifold plane we have constructed in this paper may also useful in the study of the orientifold plane in the NS5 brane background. In the Hannany-Witten setup, orientifold plane plays a crucial role in order to introduce $Sp/SO$ group and/or chiral matter contents (see {\it e.g.} \cite{Giveon:1998sr} and references therein). In some cases, to derive nonperturbative results of the gauge theory, the dynamical properties of the orientifold planes and D-branes are assumed. It would be an interesting problem to ask whether we could yield a supporting argument to such assumptions by studying the $\alpha'$ exact properties of the orientifold plane.

We can also construct nonsupersymmetric (non-BPS) orientifold planes just by throwing away the R-sector or by allowing continuous $U(1)$ charge. In some cases, they are perturbatively stable and in other cases they are not. The fate of the nonsupersymmetric orientifold plane is much less studied than the case of the non-BPS D-branes and it deserves a further study.

\subsubsection{Rolling orientifold plane?}
In \cite{Nakayama:2004yx}, we have constructed the boundary state for the rolling D-brane \cite{Kutasov:2004dj} in the NS5 brane background. This involves a nontrivial Wick-rotation of the boundary wavefunction of the Class 2 brane in the $\cN =2$ Liouville theory. Assuming this kind of Wick-rotation makes sense for other branes or crosscap states in the $\cN =2$ Liouville theory, we can construct time dependent boundary/crosscap states.

Note that this step, even if possible, is nontrivial because the direct analytic continuation in the momentum space does not work. In order to achieve the physically sensible result, we should first obtain the coordinate space wavefunction and then we perform the Wick rotation. Since the analytic continuation and Fourier transformation do not commute in general, we obtain a nontrivial momentum space wavefunction which is different from the direct analytic continuation. For the Class 2 brane, the process is schematically described as 
\begin{align}
\Psi_{\text{hairpin}}^{(\sNS)}(p,\omega) &= \frac{\Gamma(i\cQ p)\Gamma(1+i\frac{2p}{\cQ})}{\Gamma(\frac{1}{2}+\frac{\omega}{\cQ^2}+i\frac{p}{\cQ})\Gamma(\frac{1}{2}-\frac{\omega}{\cQ^2}+i\frac{p}{\cQ})} \ \to \ \text{Wick rotation:}\  Y \to it, \ \omega\to i\omega \cr \to \Psi_{\text{rolling}}^{(\sNS)}(p,i\omega) &= \frac{-i \sqrt{2}\cQ \sinh(\frac{ 2 \pi p}{\cQ})}
{2\cosh[\frac{\pi}{\cQ}(p+\frac{\omega}{\cQ})]\cosh[\frac{\pi}{\cQ}(p-\frac{\omega}{\cQ})]} \frac{\Gamma(i\cQ p)\Gamma(1+i\frac{2p}{\cQ})}{\Gamma(\frac{1}{2}+i\frac{\omega}{\cQ^2}+i\frac{p}{\cQ})\Gamma(\frac{1}{2}-i\frac{\omega}{\cQ^2}+i\frac{p}{\cQ})} \ .
\end{align}

Unfortunately, it seems difficult to complete the whole process in the case other than the Class 2 brane. One reason behind this difficulty might be that the other time dependent branes, if possible, do not seem to have a large $N$ ($\cQ \to 0$) orbit. In particular, it is difficult to construct the semiclassical orbit of the orientifold plane moving in the CHS background.\footnote{From the topological argument of section \ref{sec:top}, one can guess that the orientifold plane extends in parallel with the $\phi$ axis. After the Wick rotation, it becomes space-like orientifold from this naive view point.} Consequently, to establish the criterion about which kind of objects the Wick rotation presented in \cite{Nakayama:2004yx} can be applied to is a very important subject worth studying further to understand the general nature of the time-dependent problems in the string theory.

\subsection{Conclusion}
In this paper, we have constructed the crosscap states in the $\cN = 2$ Liouville theory from the modular bootstrap method. The crosscap states obtained in this way has some similarity with those in the bosonic or $\cN=1$ Liouville theory. We have verified our results by comparing the semiclassical limit of the crosscap wavefunction with the calculation from the minisuperspace approximation. In addition another independent consistency check based on the conformal bootstrap method has been presented. We also calculated various overlaps (M\"obius strip amplitudes) between the crosscap states and branes in the $\cN =2$ Liouville theory known in the literature. In the case of the compact branes, the overlaps with them yield the degenerate characters with the $\Omega$ insertion as expected. In the noncompact case, we have observed some peculiarities possibly caused by the IR divergence, which was also encountered in the bosonic or $\cN = 1$ Liouville theory.

Furthermore, we have discussed the topological origin of the discrete terms in the crosscap wavefunction whose derivation we have presented in Appendix \ref{sec:B}. They are related to the period integrals of the space-time theory and we have proposed that they may be calculable from the Landau-Ginzburg approach in a nontrivial dilaton background. In particular, we have found that by a suitable change of variables, we can map the problem to the Landau-Ginzburg theory with the negative power superpotential without the linear dilaton contribution, extending the known duality to the open string sector.

For a further consistency check of our results, the conformal bootstrap approach  would be important. We have shown in section \ref{sec:con} that the neutral sector of our crosscap states satisfy the conformal bootstrap equation. In the bosonic case, the conformal bootstrap approach gives a constraint on the one-point functions on the $\mathbf{RP}_2$, but does not completely determine them. It should be interesting to see what happens in the $\cN =2$ Liouville theory, extending our results in section \ref{sec:con} to the general $U(1)$ charged sector.

\section*{Acknowledgements}
The author would like to thank D.~Ghoshal, J.~Gomis, A.~Kapustin and Y.~Tachikawa for fruitful discussions and especially Y.~Sugawara for reading the manuscript and giving valuable comments to the author.
This research is supported in part by a Grant
for 21st Century COE Program ``QUESTS'' from the Ministry of Education,
Culture, Sports, Science, and Technology of Japan.
       
\appendix\sectiono{Conventions and Useful Formulae}\label{sec:A}

\subsection{Basic definitions/formulae}\label{a-3}

\begin{equation}
\Gamma(z+1) = z\Gamma(z)
\end{equation}
\begin{equation}
\Gamma(z)\Gamma(1-z) = \frac{\pi}{\sin\pi z}
\end{equation}
\begin{equation}
\Gamma(\frac{1}{2}+z)\Gamma(\frac{1}{2}-z) = \frac{\pi}{\cos\pi z}
\end{equation}
\begin{equation}
\Gamma(2z) =(2\pi)^{-1/2}2^{2z-1/2}\Gamma(z)\Gamma(z+1/2) \label{eq:ledgam}
\end{equation}
\begin{equation}
\gamma(x) \equiv \frac{\Gamma(x)}{\Gamma(1-x)}
\end{equation}
\begin{align}
F(\alpha,\beta,\gamma;z) = \frac{\Gamma(\gamma)\Gamma(\gamma-\alpha-\beta)}{\Gamma(\gamma-\alpha)\Gamma(\gamma-\beta)}F(\beta,\alpha,\alpha+\beta+1-\gamma;1-z) \cr
+ \frac{\Gamma(\gamma)\Gamma(\alpha+\beta-\gamma)}{\Gamma(\alpha)\Gamma(\beta)}(1-z)^{\gamma-\alpha-\beta}F(\gamma-\alpha,\gamma-\beta,\gamma+1-\alpha-\beta;1-z) 
\label{eq:hyperg} 
\end{align}
\begin{equation}
K_\nu(z) = \int_0^\infty e^{-z \cosh t}\cosh(\nu t)dt
\end{equation}

\subsection{Modular Functions}\label{a-4}
In this section, we review our conventions and properties of the modular functions in the $\cN = 2$ Liouville theory. Using the notation $q \equiv e^{2\pi i \tau}$ and $y \equiv e^{2\pi iz}$, we define the following theta functions with characteristics,
\begin{align}
\theta_1(\tau,z) &= i\sum_{n=-\infty}^\infty (-1)^n q^{(n-1/2)^2/2} y^{n-1/2} = 2e^{\pi i\tau/4}\sin(\pi z) \prod_{m=1}^\infty(1-q^m)(1-yq^m)(1-y^{-1}q^m) \cr\theta_2(\tau,z) &= \sum_{n=-\infty}^\infty q^{(n-1/2)^2/2} y^{n-1/2} = 2e^{\pi i\tau/4}\cos(\pi z) \prod_{m=1}^\infty(1-q^m)(1+yq^m)(1+y^{-1}q^m) \cr
\theta_3(\tau,z) &= \sum_{n=-\infty}^\infty q^{n^2/2}y^n = \prod_{m=1}^\infty(1-q^m)(1+yq^{m-1/2})(1+y^{-1}q^{m-1/2}) \cr
\theta_4(\tau,z) &= \sum_{n=-\infty}^\infty (-1)^n q^{n^2/2} y^{n} =\prod_{m=1}^\infty(1-q^m)(1-yq^{m-1/2})(1-y^{-1}q^{m-1/2}) \ ,
\end{align}
and when we omit the second argument $z$, we always mean $z=0$. The Dedekind $\eta$ function is defined as
\begin{equation}
\eta(\tau) = q^{\frac{1}{24}}\prod_{n=1}^\infty(1-q^n) \ .
\end{equation}

The modular transformation is given by
\begin{align}
\theta_3(\tau + 1,z) &= \theta_4(\tau,z) \cr
\theta_4(\tau + 1,z) &= \theta_3(\tau,z) \cr
\theta_2(\tau + 1,z) &= e^{\pi i/4} \theta_2(\tau,z) \cr
\theta_1(\tau + 1,z) &= e^{\pi i/4} \theta_1(\tau,z) \cr
\eta(\tau + 1) &= e^{\pi i/12} \eta(\tau) \label{eq:mot}
\end{align}
for the T-transformation, and 
\begin{align}
\theta_3(-1/\tau,z/\tau) &= \sqrt{-i\tau}\exp(\pi iz^2/\tau) \theta_3(\tau,z) \cr
\theta_4(-1/\tau,z/\tau) &= \sqrt{-i\tau}\exp(\pi iz^2/\tau) \theta_2(\tau,z) \cr
\theta_2(-1/\tau,z/\tau) &= \sqrt{-i\tau}\exp(\pi iz^2/\tau) \theta_4(\tau,z) \cr
\theta_1(-1/\tau,z/\tau) &= -i\sqrt{-i\tau}\exp(\pi iz^2/\tau) \theta_1(\tau,z) \cr
\eta(-1/\tau) &= \sqrt{-i\tau} \eta(\tau)\ , \label{eq:mos}
\end{align}
for the S-transformation.
Therefore the modular transformation for the $\cN =2$ massive character is written as
\begin{align}
\mathrm{ch}^{(\sNS)}(p,\omega;-\frac{1}{\tau},\frac{z}{\tau}) = e^{i\pi\frac{\hat{c}z^2}{\tau}}\frac{2}{\cQ} \int_0^\infty dp' \int_{-\infty}^{\infty}d\omega' e^{-2\pi i\frac{\omega\omega'}{\cQ^2}} \cos(2\pi pp') \mathrm{ch}^{(\sNS)}(p',\omega';\tau,z) \ . \label{modum}
\end{align}

For the application to the unoriented Liouville theory, the following modular transformation is needed (we will set $z=0$),
\begin{align}
\eta\left(\frac{i}{4t} + \frac{1}{2}\right) &= \sqrt{2t}\eta\left(it + \frac{1}{2}\right) \cr
\theta_3\left(\frac{i}{4t} + \frac{1}{2}\right) &= \sqrt{2t}e^{\frac{\pi i}{4}}\theta_4\left(it + \frac{1}{2}\right) \cr
\theta_4\left(\frac{i}{4t} + \frac{1}{2}\right) &= \sqrt{2t}e^{\frac{-\pi i}{4}}\theta_3\left(it + \frac{1}{2}\right). \label{eq:modun}
\end{align}
To obtain these, we have to perform the succeeding modular transformations $TSTTS$.\footnote{Since, we are replacing $\tau \to \tau +\frac{1}{2}$ only in the oscillator part, we are actually dealing with the ``hatted" characters (see {\it e.g.} \cite{Angelantonj:2002ct} and references therein). In this case, the modular transformation is given by $P=T^{1/2}STTST^{1/2}$, but this does not make any difference in our calculation.} We can show this explicitly by setting $t_4 = \frac{i}{4t} + \frac{1}{2}$, then
\begin{align}
t_3 &= -\frac{1}{t_4} = -\frac{2it}{it-\frac{1}{2}} \cr
t_2 &= t_3 + 2 = -\frac{1}{it-\frac{1}{2}} \cr
t_1 &= -\frac{1}{t_2} = it - \frac{1}{2} \cr
t_0 &= t_1 + 1 = it + \frac{1}{2}\ .
\end{align}
Therefore, the modular transformation becomes (we take $\eta$ function for example)
\begin{align}
\eta(t_4) &= \eta (-1/t_3) \cr
	  &= \sqrt{-it_3}\eta(t_3) \cr
	&= \sqrt{-i(t_2-2)}\eta(t_2-2) \cr
	&= \sqrt{-i(t_2-2)}e^{-\frac{i\pi}{6}}\eta(t_2) \cr
	&= \sqrt{-i(-\frac{1}{t_1} - 2)}e^{-\frac{i\pi}{6}} \sqrt{-it_1}\eta(t_1) \cr
	&= \sqrt{1+2t_1} e^{-\frac{i\pi}{6}}\eta(t_1)\cr
	&= \sqrt{2t_0 -1} e^{-\frac{i\pi}{4}}\eta(t_0) \cr
	&= \sqrt{2t} \eta(it + \frac{1}{2})\ .
\end{align}
Similarly we can obtain the transformation for $\theta$ functions.

Then we can derive the modular transformation of the massive character with the $\Omega$ insertion (the M\"obius strip amplitude):
\begin{align}
\mathrm{ch}_{\Omega}^{(\sNS)}(p,\omega;-\frac{1}{4\tau}) = \frac{i}{\cQ} \int_0^\infty dp' \int_{-\infty}^{\infty}d\omega' e^{-\pi i\frac{\omega\omega'}{\cQ^2}} \cos(\pi pp') \mathrm{ch}_{\widetilde{\Omega}}^{(\sNS)}(p',\omega';\tau) \ .
\end{align}
Note that the argument of the integrand $e^{-\pi i\frac{\omega\omega'}{\cQ^2}} \cos(\pi pp') $ is half that of the cylinder case \eqref{modum}, which is typical of the M\"obius strip amplitude.

Let us finally list the general channel duality properties of the Ishibashi states we use in the main text. The easiest way to derive the following results is to calculate directly those quantities by using the free field Ishibashi states. The left hand side is the tree exchange channel and the right hand side is the loop channel which can be obtained by the modular transformation: $t =1/T$ for cylinder $t=1/2T$ for the Klein bottle and $t=1/4T$ for the M\"obius strip.

\begin{align}
\langle B \pm| e^{-\pi TH^{(c)}} |B \pm\rangle_{\sR\sR} &\to \NS; \ (-1)^f \cr
\langle B \pm| e^{-\pi TH^{(c)}} |B \mp\rangle_{\sR\sR} &\to \R; \ (-1)^f \cr
\langle B \pm| e^{-\pi TH^{(c)}} |B \pm\rangle_{\sNS\sNS} &\to \NS; \ 1 \cr
\langle B \pm| e^{-\pi TH^{(c)}} |B \mp\rangle_{\sNS\sNS} &\to \R; \ 1 
\end{align}
for the cylinder.
\begin{align}
\langle C \pm| e^{-\pi TH^{(c)}} |C \pm\rangle_{\sR\sR} &\to \NS-\NS; \ ((-1)^f+(-1)^{\bar{f}})\cdot\Omega \cr
\langle C \pm| e^{-\pi TH^{(c)}} |C \mp\rangle_{\sR\sR} &\to \R-\R; \ ((-1)^f+(-1)^{\bar{f}})\cdot\Omega \cr
\langle C \pm| e^{-\pi TH^{(c)}} |C \pm\rangle_{\sNS\sNS} &\to \NS-\NS; \ (1+(-1)^{f+\bar{f}})\cdot\Omega \cr
\langle C \pm| e^{-\pi TH^{(c)}} |C \mp\rangle_{\sNS\sNS} &\to \R-\R; \ (1+(-1)^{f+\bar{f}})\cdot\Omega 
\end{align}
for the Klein bottle. 
\begin{align}
\langle B \pm| e^{-\pi TH^{(c)}} |C \pm\rangle_{\sR\sR} &\to \R; \ (-1)^f\cdot\Omega \cr
\langle B \pm| e^{-\pi TH^{(c)}} |C \mp\rangle_{\sR\sR} &\to \R; \ 1 \cdot\Omega \cr
\langle B \pm| e^{-\pi TH^{(c)}} |C \pm\rangle_{\sNS\sNS} &\to \NS; \ (-1)^f \cdot\Omega \cr
\langle B \pm| e^{-\pi TH^{(c)}} |C \mp\rangle_{\sNS\sNS} &\to \NS; \ 1 \cdot \Omega
\end{align}
for the M\"obius strip (up to a phase factor).

\sectiono{Modular Transformation of $\Omega$ Inserted Character}\label{sec:B}
In this Appendix, we derive the modular transformation of the character for the graviton representation with the $\Omega$ insertion following the prescription by \cite{Miki:1989ri}. To do this, we only need to know the modular transformation of the (spectral-flowed) massless characters because there is a relation:
\begin{equation}
\chi^{(\sNS)}_G(r=0;\tau) = \chi^{(\sNS)}(h=0,j=0;\tau) - \chi^{(\sNS)}_M(r=0,s=N;\tau) - \chi^{(\sNS)}_M(r=-1,s=2K;\tau) \ .\label{key2}
\end{equation}
Similarly, we have 
\begin{align}
\chi^{(\sNS)}_{G,\Omega}(r=0;\tau) &= \chi^{(\sNS)}_{\Omega}(h=j=0;\tau) - i\chi^{(\sNS)}_{M,\Omega}(r=0,s=N;\tau) - \chi^{(\sNS)}_{M,\Omega'}(r=-1,s=2K;\tau) \cr 
&= \chi^{(\sNS)}_{\Omega}(h=j=0;\tau) -2i \chi^{(\sNS)}_{M,\Omega}(r=0,s=N;\tau) 
\ ,\label{key}
\end{align}
where $\chi^{(\sNS)}_{M,\Omega'}(r,s;\tau)$ is given by replacing $i$ in the denominator of \eqref{masc} with $-i$:
\begin{equation}
\chi_{M,\Omega'}^{(\sNS)}(r,s;\tau) = \sum_{m\in \bsz} \frac{q^{(s-K)(m+\frac{2r+1}{2N})}}{1-iq^{N(m+\frac{2r+1}{2N})}}q^{NK(m+\frac{2r+1}{2N})^2}e^{\frac{\pi i}{8}}\frac{\theta_3(\tau+\frac{1}{2})}{\eta(\tau+\frac{1}{2})^3} \ .\label{masc2}
\end{equation}

We first define 
\begin{equation}
I_{\Omega^{\pm}} (k,a,b;\tau) = \sum_{r\in \bsz +\frac{1}{2}} e^{2\pi i ar} \frac{q^{rb}}{1\pm iq^{r}}q^{\frac{k}{2}r^2} \ .
\end{equation}
Then from Cauchy's theorem,
\begin{equation}
\frac{i}{\tau} I_{\Omega^\pm} (k,a,b;-\frac{1}{4\tau}) = \frac{1}{2\pi i} \left[\int_{-\infty-i\epsilon}^{\infty-i\epsilon}-\int_{-\infty+i\epsilon}^{\infty+i\epsilon}\right] dx \frac{4i\pi e^{-2\pi(b-\frac{1}{2})x + 8\pi it x(a-\frac{1}{2})-4 k\pi t x^2}}{\cos[4\pi tx] (e^{\pi x} \pm ie^{-\pi x})} \ , \label{uno}
\end{equation}
where $\tau = it$. Next, we use the expansion $\frac{1}{2\cos(\pi t x)} = \sum_{n\ge 0} (-1)^n e^{\pm(2n+1)i\pi t} $ for $\pm \mathrm{Im} x >0$ respectively to rewrite  \eqref{uno} as (we consider $\Omega^+$ case)
\begin{equation}
\sum_{r\in \bsz +a} \frac{1}{2\pi i} \int dx J^b(r,x) \ ,
\end{equation}
where 
\begin{equation}
J^b(r,x) = \frac{8i\pi e^{i \pi (r-a) + 2\pi i(4t) rx - 2\pi (b-\frac{1}{2})x - 4k \pi t x^2}}{e^{\pi x} + ie^{-\pi x}} \ .
\end{equation}
Modifying the contour of $x$, we divide the contribution in two parts:
\begin{equation}
\sum_{r\in \bsz +a} \frac{1}{2\pi i} \int dx J^b(r,x) = J_1^{a,b} + J_2^{a,b} \ .
\end{equation}
The continuum part $J_2^{a,b}$ is given by 
\begin{align}
J_2^{a,b} &= \sum_{r\in \bsz + a} \frac{1}{2\pi i} \int dx \frac{8i\pi e^{i\pi(r-a)-2\pi b (x+\frac{ir}{k})-4k\pi t x^2 -\frac{4\pi t r^2}{k}}}{1+ie^{-2\pi(x+\frac{ir}{k})}} \cr
&= \sum_{r\in \bsz + a} \frac{1}{2\pi i} \int dp \frac{4\pi i}{\sqrt{k}}\frac{e^{i\pi (r-a)} q^{\frac{2r^2}{k}} e^{-2\pi b(\frac{p}{2\sqrt{k}}+\frac{ir}{k})}}{1+ie^{-2\pi(\frac{p}{2\sqrt{k}}+\frac{ir}{k})}} q^{\frac{p^2}{2}} \ ,
\end{align}
where the integration should be understood as the principal value.

On the other hand, the discrete part $J_1^{a,b}$ is given by the contribution from the pole at $x = is$ where $s \in \bz + \frac{3}{4}$ (in the $\Omega^-$ case, $s \in \bz + \frac{1}{4}$) as
\begin{equation}
J_1^{a,b} = \left(\sum_{r>0} \sum_{\frac{r}{k}>s>0} - \sum_{r<0}\sum_{\frac{r}{k}<s<0} \right)j^b(r,s) \ ,
\end{equation}
where the summation is taken over $r\in \bz + a$ and $j(r,s)$ is the residue of $J^b(r,x)$ at $x = is$ with $s\in \bz+\frac{3}{4}$.\footnote{Strictly speaking, if $r = ks$, the contribution from the residue should be half of the other part \cite{Miki:1989ri}. However, this does not make any difference once we take the continuum limit, so we will neglect this boundary contribution (see also \cite{Eguchi:2003ik}). } Then we take the summation over $r$ first
\begin{equation}
J_1^{a,b} = \left(\sum_{s>0} \sum_{r>0} - \sum_{s<0}\sum_{r<0} \right)j^b(ks+r+\delta ,s) \ ,
\end{equation}
where $r\in {\bf Z}$ and $ 0 \le \delta(a,s) <1$ is defined as $\delta(a,s) \equiv a- ks $ (mod ${\bf Z}$).

Noticing 
\begin{equation}
j(ks+r+\delta) = 8ie^{-i\pi a} (e^{i\pi} q^{4s})^{r+\delta} e^{2\pi i(\frac{k}{2}-b)s} q^{2ks^2} \ ,
\end{equation}
we obtain 
\begin{equation}
J_1^{a,b} = 8i\sum_{s\in \bsz+\frac{3}{4}} e^{-i\pi(a-\delta)} \frac{q^{4s\delta}}{1+q^{4s}} q^{2ks^2} e^{2\pi i (\frac{k}{2}-b)s} \ . \label{due}
\end{equation}

Now we would like to obtain the modular transformation of the character for the massless representation with the $\Omega$ insertion. Firstly, we note that
\begin{equation}
\chi_{M,\Omega}^{(\sNS)}(r,s;\tau) = \frac{1}{N} \sum_{j=0}^{N-1} e^{-2\pi \frac{(2r+1)}{2N} j} I_{\Omega^+} (\frac{2K}{N}, \frac{j}{N},\frac{s-K}{N},\tau) e^{\frac{i\pi}{8}} \frac{\theta_3(\tau+\frac{1}{2})}{\eta(\tau +\frac{1}{2})^3} \ , \label{tri}
\end{equation}
and 
\begin{equation}
\chi_{M,\Omega'}^{(\sNS)}(r,s;\tau) = \frac{1}{N} \sum_{j=0}^{N-1} e^{-2\pi \frac{(2r+1)}{2N} j} I_{\Omega^-} (\frac{2K}{N}, \frac{j}{N},\frac{s-K}{N},\tau) e^{\frac{i\pi}{8}} \frac{\theta_3(\tau+\frac{1}{2})}{\eta(\tau +\frac{1}{2})^3} \ . \label{tri2} 
\end{equation}
By using \eqref{key} and \eqref{uno}, we obtain contribution for the continuum part from the massless representation
\begin{equation}
-2i\frac{1}{N}\sum_{j=0}^{N-1} \sum_{r'\in \bsz + \frac{j}{N}} \frac{1}{2\pi i} \int dp \frac{-2\pi}{\cQ}\frac{e^{\pi(i(r'-\frac{2j}{N})-2\pi(\frac{s-K}{N})(\frac{p}{2\cQ}+\frac{ir'}{\cQ^2})}}{1+ie^{-\pi(\frac{p}{\cQ}+\frac{2ir'}{\cQ^2})}}q^{\frac{{r'}^2}{2}+\frac{p^2}{2}} \mathrm{ch}_{\widetilde\Omega}^{(\sNS)}(p,r';\tau) \ .
\end{equation}
When we take the continuum limit, it becomes
\begin{equation}
= -\frac{i}{\cQ} \int_{-\infty}^{\infty} d\omega\int_{-\infty}^{\infty}dp e^{\frac{\pi p \cQ}{2}} \frac{1}{1-ie^{\pi(\frac{p}{\cQ}+i\frac{\omega}{\cQ^2})}} \mathrm{ch}_{\widetilde \Omega}^{(\sNS)}(p,\omega;\tau)\ .
\end{equation}
To connect this result to what we have used in the main text \eqref{moduo}, notice that we can symmetrize the integrand as $\omega \to -\omega$ and $p \to -p$.
Then we can use the identity
\begin{align}
-e^{-\frac{\pi p \cQ}{2}} \left(\frac{1}{1+ie^{\pi (\frac{p}{\cQ} + i\frac{\omega}{\cQ^2})}} + \frac{1}{1+ie^{\pi (\frac{p}{\cQ} - i\frac{\omega}{\cQ^2})}} \right)- e^{\frac{\pi p \cQ}{2}} \left( \frac{1}{1+ie^{\pi (-\frac{p}{\cQ} + i\frac{\omega}{\cQ^2})}} +\frac{1}{1+ie^{\pi (-\frac{p}{\cQ} - i\frac{\omega}{\cQ^2})}}\right)  \cr
= e^{\frac{\pi p \cQ}{2}} \left(\frac{1}{1-ie^{\pi (\frac{p}{\cQ} + i\frac{\omega}{\cQ^2})}} +\frac{1}{1-ie^{\pi (\frac{p}{\cQ} -i\frac{\omega}{\cQ^2})}} \right)+ e^{-\frac{\pi p \cQ}{2}} \left(\frac{1}{1-ie^{\pi (-\frac{p}{\cQ} + i\frac{\omega}{\cQ^2})}} + \frac{1}{1-ie^{\pi (-\frac{p}{\cQ} - i\frac{\omega}{\cQ^2})}} \right)\ ,
\end{align}
which reproduces \eqref{moduo} (after adding the contribution from the massive part in \eqref{key}).

Now let us move on to the discrete part. From \eqref{tri} and \eqref{due}, the discrete part is given by 
\begin{align}
 \chi_{M,\Omega}^{(\sNS)}(r,s;-\frac{1}{4\tau})|_{disc} 
&= -4 \frac{1}{N}\sum_{s'=K+1}^{N+K-1}\sum_{r'\in \bsz_N}\sum_{m\in \bsz} e^{2\pi i\frac{(s+2Kr)(s'+2Kr')-(s-K)(s'-K)-(\frac{s}{2}+rK)K}{2NK}} \cr 
&\times \frac{q^{[3+ (4Nm+4r')]\frac{s'-K}{N}}}{1+q^{3+(4Nm+4r')}} q^{\frac{4K}{N}(Nm+r'+\frac{3}{4})^2}e^{-\frac{\pi i}{8}}\frac{\theta_4(\tau+\frac{1}{2})}{\eta(\tau+\frac{1}{2})^3} \ .
\end{align}

In the continuum limit, the $\Omega^{+}$ contribution of \eqref{key} becomes
\begin{equation}
-4 \sum_{r'\in \bsz} \int_{\frac{\cQ^2}{2}}^{1+\frac{\cQ^2}{2}} d\lambda e^{i\pi(\lambda-\frac{\cQ^2}{2})} \frac{q^{(3+4r')(\lambda-\frac{\cQ^2}{2})}}{1+q^{3+4r'}} q^{\frac{\cQ^2}{8}(4r'+3)^2}e^{-\frac{\pi i}{8}}\frac{\theta_4(\tau+\frac{1}{2})}{\eta(\tau+\frac{1}{2})^3} \ .
\end{equation} 
On the other hand, from the $\Omega^{-}$ contribution of \eqref{key}, we also have
\begin{equation}
-4 \sum_{r'\in \bsz} \int_{\frac{\cQ^2}{2}}^{1+\frac{\cQ^2}{2}} d\lambda e^{i\pi(\lambda-\frac{\cQ^2}{2})} \frac{q^{(1+4r')(\lambda-\frac{\cQ^2}{2})}}{1+q^{1+4r'}} q^{\frac{\cQ^2}{8}(4r'+1)^2}e^{-\frac{\pi i}{8}}\frac{\theta_4(\tau+\frac{1}{2})}{\eta(\tau+\frac{1}{2})^3} \ ,
\end{equation} 
therefore we totally obtain
\begin{equation}
-4 \sum_{r'\in \bsz} \int_{\frac{\cQ^2}{2}}^{1+\frac{\cQ^2}{2}} d\lambda e^{i\pi(\lambda-\frac{\cQ^2}{2})} \frac{q^{(1+2r')(\lambda-\frac{\cQ^2}{2})}}{1+q^{1+2r'}} q^{\frac{\cQ^2}{8}(2r'+1)^2}e^{-\frac{\pi i}{8}}\frac{\theta_4(\tau+\frac{1}{2})}{\eta(\tau+\frac{1}{2})^3} \ ,
\end{equation} 
To go further, we realize that $2/(1+q) = 1/(1+iq^{1/2}) + 1/(1-iq^{1/2})$, and introduce $\omega' = 2\lambda - \cQ^2$. Then
\begin{equation}
\mathrm{ch}_{G,\Omega}^{(\sNS)}(-\frac{1}{4\tau})|_{disc} = i\sum_{r'\in \bsz} \int_0^2 d\omega' e^{i\pi \frac{\omega'}{2}} \left(\frac{q^{(r'+\frac{1}{2})\omega'}}{1+iq^{r'+\frac{1}{2}}} + \frac{q^{(r'+\frac{1}{2})\omega'}}{1-iq^{r'+\frac{1}{2}}}\right)q^{\frac{\cQ^2}{8}(2r'+1)^2} e^{-\frac{\pi i}{8}}\frac{\theta_4(\tau+\frac{1}{2})}{\eta(\tau+\frac{1}{2})^3} \ .\label{disp}
\end{equation}
Note that the second term is the $\widetilde{\Omega}$ counterpart of \eqref{quattro}.\footnote{However, the first term contains wrongly projected characters, so in the full theory, we hope these terms are GSO projected out.} It is interesting to compare it with the discrete terms for the modular transformation of the cylinder amplitude for the identity (graviton) representation
\begin{equation}
\mathrm{ch}_{G}^{(\sNS)}(-\frac{1}{\tau})|_{disc} = \sum_{r'\in\bsz} \int_0^1 d\omega' 2 e^{i\pi \omega' }\frac{q^{(r'+\frac{1}{2})\omega'}}{1+q^{r'+\frac{1}{2}}} q^{\frac{\cQ^2}{8}(2r'+1)^2} \frac{\theta_3(\tau)}{\eta(\tau)^3} \ .\label{disc}
\end{equation}

\bibliographystyle{utcaps}
\bibliography{crosscap}

\providecommand{\href}[2]{#2}\begingroup\raggedright\begin{thebibliography}{10}

\bibitem{Teschner:2001rv}
J.~Teschner, ``Liouville theory revisited,'' {\em Class. Quant. Grav.} {\bf 18}
  (2001) R153--R222,
\href{http://www.arXiv.org/abs/hep-th/0104158}{{\tt hep-th/0104158}}.

\bibitem{Nakayama:2004vk}
Y.~Nakayama, ``Liouville field theory: A decade after the revolution,'' {\em
  Int. J. Mod. Phys.} {\bf A19} (2004) 2771--2930,
\href{http://www.arXiv.org/abs/hep-th/0402009}{{\tt hep-th/0402009}}.

\bibitem{McGreevy:2003kb}
J.~McGreevy and H.~Verlinde, ``Strings from tachyons: The c = 1 matrix
  reloaded,'' {\em JHEP} {\bf 12} (2003) 054,
\href{http://www.arXiv.org/abs/hep-th/0304224}{{\tt hep-th/0304224}}.

\bibitem{Takayanagi:2003sm}
T.~Takayanagi and N.~Toumbas, ``A matrix model dual of type 0B string theory in
  two dimensions,'' {\em JHEP} {\bf 07} (2003) 064,
\href{http://www.arXiv.org/abs/hep-th/0307083}{{\tt hep-th/0307083}}.

\bibitem{Douglas:2003up}
M.~R. Douglas {\em et al.}, ``A new hat for the c = 1 matrix model,''
\href{http://www.arXiv.org/abs/hep-th/0307195}{{\tt hep-th/0307195}}.

\bibitem{Distler:1989nt}
J.~Distler, Z.~Hlousek, and H.~Kawai, ``SUPERLIOUVILLE THEORY AS A
  TWO-DIMENSIONAL, SUPERCONFORMAL SUPERGRAVITY THEORY,'' {\em Int. J. Mod.
  Phys.} {\bf A5} (1990)
391.

\bibitem{Kutasov:1990ua}
D.~Kutasov and N.~Seiberg, ``Noncritical superstrings,'' {\em Phys. Lett.} {\bf
  B251} (1990)
67--72.

\bibitem{McGreevy:2003dn}
J.~McGreevy, S.~Murthy, and H.~Verlinde, ``Two-dimensional superstrings and the
  supersymmetric matrix model,'' {\em JHEP} {\bf 04} (2004) 015,
\href{http://www.arXiv.org/abs/hep-th/0308105}{{\tt hep-th/0308105}}.

\bibitem{Verlinde:2004gt}
H.~Verlinde, ``Superstrings on AdS(2) and superconformal matrix quantum
  mechanics,''
\href{http://www.arXiv.org/abs/hep-th/0403024}{{\tt hep-th/0403024}}.

\bibitem{Callan:1991at}
C.~G.~. Callan, J.~A. Harvey, and A.~Strominger, ``Supersymmetric string
  solitons,''
\href{http://www.arXiv.org/abs/hep-th/9112030}{{\tt hep-th/9112030}}.

\bibitem{Fateev:2000ik}
V.~Fateev, A.~B. Zamolodchikov, and A.~B. Zamolodchikov, ``Boundary Liouville
  field theory. I: Boundary state and boundary two-point function,''
\href{http://www.arXiv.org/abs/hep-th/0001012}{{\tt hep-th/0001012}}.

\bibitem{Teschner:2000md}
J.~Teschner, ``Remarks on Liouville theory with boundary,''
\href{http://www.arXiv.org/abs/hep-th/0009138}{{\tt hep-th/0009138}}.

\bibitem{Zamolodchikov:2001ah}
A.~B. Zamolodchikov and A.~B. Zamolodchikov, ``Liouville field theory on a
  pseudosphere,''
\href{http://www.arXiv.org/abs/hep-th/0101152}{{\tt hep-th/0101152}}.

\bibitem{Fukuda:2002bv}
T.~Fukuda and K.~Hosomichi, ``Super Liouville theory with boundary,'' {\em
  Nucl. Phys.} {\bf B635} (2002) 215--254,
\href{http://www.arXiv.org/abs/hep-th/0202032}{{\tt hep-th/0202032}}.

\bibitem{Ahn:2002ev}
C.~Ahn, C.~Rim, and M.~Stanishkov, ``Exact one-point function of N = 1
  super-Liouville theory with boundary,'' {\em Nucl. Phys.} {\bf B636} (2002)
  497--513,
\href{http://www.arXiv.org/abs/hep-th/0202043}{{\tt hep-th/0202043}}.

\bibitem{Hikida:2002bt}
Y.~Hikida, ``Liouville field theory on a unoriented surface,'' {\em JHEP} {\bf
  05} (2003) 002,
\href{http://www.arXiv.org/abs/hep-th/0210305}{{\tt hep-th/0210305}}.

\bibitem{Nakayama:2003ep}
Y.~Nakayama, ``Tadpole cancellation in unoriented Liouville theory,'' {\em
  JHEP} {\bf 11} (2003) 017,
\href{http://www.arXiv.org/abs/hep-th/0309063}{{\tt hep-th/0309063}}.

\bibitem{Gomis:2003vi}
J.~Gomis and A.~Kapustin, ``Two-dimensional unoriented strings and matrix
  models,'' {\em JHEP} {\bf 06} (2004) 002,
\href{http://www.arXiv.org/abs/hep-th/0310195}{{\tt hep-th/0310195}}.

\bibitem{Bergman:2003yp}
O.~Bergman and S.~Hirano, ``The cap in the hat: Unoriented 2D strings and
  matrix(- vector) models,'' {\em JHEP} {\bf 01} (2004) 043,
\href{http://www.arXiv.org/abs/hep-th/0311068}{{\tt hep-th/0311068}}.

\bibitem{Eguchi:2003ik}
T.~Eguchi and Y.~Sugawara, ``Modular bootstrap for boundary N = 2 Liouville
  theory,'' {\em JHEP} {\bf 01} (2004) 025,
\href{http://www.arXiv.org/abs/hep-th/0311141}{{\tt hep-th/0311141}}.

\bibitem{Ahn:2003tt}
C.~Ahn, M.~Stanishkov, and M.~Yamamoto, ``One-point functions of N = 2
  super-Liouville theory with boundary,'' {\em Nucl. Phys.} {\bf B683} (2004)
  177--195,
\href{http://www.arXiv.org/abs/hep-th/0311169}{{\tt hep-th/0311169}}.

\bibitem{Ahn:2004qb}
C.~Ahn, M.~Stanishkov, and M.~Yamamoto, ``ZZ-branes of N = 2 super-Liouville
  theory,'' {\em JHEP} {\bf 07} (2004) 057,
\href{http://www.arXiv.org/abs/hep-th/0405274}{{\tt hep-th/0405274}}.

\bibitem{Hosomichi:2004ph}
K.~Hosomichi, ``N=2 Liouville Theory with Boundary,''
\href{http://www.arXiv.org/abs/hep-th/0408172}{{\tt hep-th/0408172}}.

\bibitem{Lewellen:1991tb}
D.~C. Lewellen, ``Sewing constraints for conformal field theories on surfaces
  with boundaries,'' {\em Nucl. Phys.} {\bf B372} (1992)
654--682.

\bibitem{Fioravanti:1993hf}
D.~Fioravanti, G.~Pradisi, and A.~Sagnotti, ``Sewing constraints and
  nonorientable open strings,'' {\em Phys. Lett.} {\bf B321} (1994) 349--354,
\href{http://www.arXiv.org/abs/hep-th/9311183}{{\tt hep-th/9311183}}.

\bibitem{Cardy:1989ir}
J.~L. Cardy, ``BOUNDARY CONDITIONS, FUSION RULES AND THE VERLINDE FORMULA,''
  {\em Nucl. Phys.} {\bf B324} (1989)
581.

\bibitem{Hori:2001ax}
K.~Hori and A.~Kapustin, ``Duality of the fermionic 2d black hole and N = 2
  Liouville theory as mirror symmetry,'' {\em JHEP} {\bf 08} (2001) 045,
\href{http://www.arXiv.org/abs/hep-th/0104202}{{\tt hep-th/0104202}}.

\bibitem{Ribault:2003ss}
S.~Ribault and V.~Schomerus, ``Branes in the 2-D black hole,'' {\em JHEP} {\bf
  02} (2004) 019,
\href{http://www.arXiv.org/abs/hep-th/0310024}{{\tt hep-th/0310024}}.

\bibitem{Israel:2004jt}
D.~Israel, A.~Pakman, and J.~Troost, ``D-branes in N = 2 Liouville theory and
  its mirror,''
\href{http://www.arXiv.org/abs/hep-th/0405259}{{\tt hep-th/0405259}}.

\bibitem{Fotopoulos:2004ut}
A.~Fotopoulos, V.~Niarchos, and N.~Prezas, ``D-branes and extended characters
  in SL(2,R)/U(1),''
\href{http://www.arXiv.org/abs/hep-th/0406017}{{\tt hep-th/0406017}}.

\bibitem{Ahn:2002sx}
C.~Ahn, C.~Kim, C.~Rim, and M.~Stanishkov, ``Duality in N = 2 super-Liouville
  theory,'' {\em Phys. Rev.} {\bf D69} (2004) 106011,
\href{http://www.arXiv.org/abs/hep-th/0210208}{{\tt hep-th/0210208}}.

\bibitem{Miki:1989ri}
K.~Miki, ``THE REPRESENTATION THEORY OF THE SO(3) INVARIANT SUPERCONFORMAL
  ALGEBRA,'' {\em Int. J. Mod. Phys.} {\bf A5} (1990)
1293.

\bibitem{Ishibashi:1988kg}
N.~Ishibashi, ``THE BOUNDARY AND CROSSCAP STATES IN CONFORMAL FIELD THEORIES,''
  {\em Mod. Phys. Lett.} {\bf A4} (1989)
251.

\bibitem{Eguchi:2004yi}
T.~Eguchi and Y.~Sugawara, ``SL(2,R)/U(1) supercoset and elliptic genera of
  non-compact Calabi-Yau manifolds,'' {\em JHEP} {\bf 05} (2004) 014,
\href{http://www.arXiv.org/abs/hep-th/0403193}{{\tt hep-th/0403193}}.

\bibitem{Baseilhac:1998eq}
P.~Baseilhac and V.~A. Fateev, ``Expectation values of local fields for a
  two-parameter family of integrable models and related perturbed conformal
  field theories,'' {\em Nucl. Phys.} {\bf B532} (1998) 567--587,
\href{http://www.arXiv.org/abs/hep-th/9906010}{{\tt hep-th/9906010}}.

\bibitem{Vafa:1990mu}
C.~Vafa, ``Topological Landau-Ginzburg models,'' {\em Mod. Phys. Lett.} {\bf
  A6} (1991)
337--346.

\bibitem{Hori:2000kt}
K.~Hori and C.~Vafa, ``Mirror symmetry,''
\href{http://www.arXiv.org/abs/hep-th/0002222}{{\tt hep-th/0002222}}.

\bibitem{Hori:2000ck}
K.~Hori, A.~Iqbal, and C.~Vafa, ``D-branes and mirror symmetry,''
\href{http://www.arXiv.org/abs/hep-th/0005247}{{\tt hep-th/0005247}}.

\bibitem{Brunner:2003zm}
I.~Brunner and K.~Hori, ``Orientifolds and mirror symmetry,''
\href{http://www.arXiv.org/abs/hep-th/0303135}{{\tt hep-th/0303135}}.

\bibitem{Hori:2002cd}
K.~Hori and A.~Kapustin, ``Worldsheet descriptions of wrapped NS five-branes,''
  {\em JHEP} {\bf 11} (2002) 038,
\href{http://www.arXiv.org/abs/hep-th/0203147}{{\tt hep-th/0203147}}.

\bibitem{Klemm:1996bj}
A.~Klemm, W.~Lerche, P.~Mayr, C.~Vafa, and N.~P. Warner, ``Self-Dual Strings
  and N=2 Supersymmetric Field Theory,'' {\em Nucl. Phys.} {\bf B477} (1996)
  746--766,
\href{http://www.arXiv.org/abs/hep-th/9604034}{{\tt hep-th/9604034}}.

\bibitem{Lerche:2000uy}
W.~Lerche, ``On a boundary CFT description of nonperturbative N = 2 Yang-Mills
  theory,''
\href{http://www.arXiv.org/abs/hep-th/0006100}{{\tt hep-th/0006100}}.

\bibitem{Kapustin:2003ga}
A.~Kapustin and Y.~Li, ``Topological correlators in Landau-Ginzburg models with
  boundaries,'' {\em Adv. Theor. Math. Phys.} {\bf 7} (2004) 727--749,
\href{http://www.arXiv.org/abs/hep-th/0305136}{{\tt hep-th/0305136}}.

\bibitem{Kapustin:2003rc}
A.~Kapustin and Y.~Li, ``D-branes in topological minimal models: The
  Landau-Ginzburg approach,'' {\em JHEP} {\bf 07} (2004) 045,
\href{http://www.arXiv.org/abs/hep-th/0306001}{{\tt hep-th/0306001}}.

\bibitem{Brunner:2003dc}
I.~Brunner, M.~Herbst, W.~Lerche, and B.~Scheuner, ``Landau-Ginzburg
  realization of open string TFT,''
\href{http://www.arXiv.org/abs/hep-th/0305133}{{\tt hep-th/0305133}}.

\bibitem{Herbst:2004ax}
M.~Herbst and C.-I. Lazaroiu, ``Localization and traces in open-closed
  topological Landau- Ginzburg models,''
\href{http://www.arXiv.org/abs/hep-th/0404184}{{\tt hep-th/0404184}}.

\bibitem{Herbst:2004zm}
M.~Herbst, C.-I. Lazaroiu, and W.~Lerche, ``D-brane effective action and
  tachyon condensation in topological minimal models,''
\href{http://www.arXiv.org/abs/hep-th/0405138}{{\tt hep-th/0405138}}.

\bibitem{Lukyanov:2003nj}
S.~L. Lukyanov, E.~S. Vitchev, and A.~B. Zamolodchikov, ``Integrable model of
  boundary interaction: The paperclip,'' {\em Nucl. Phys.} {\bf B683} (2004)
  423--454,
\href{http://www.arXiv.org/abs/hep-th/0312168}{{\tt hep-th/0312168}}.

\bibitem{Nakayama:2004yx}
Y.~Nakayama, Y.~Sugawara, and H.~Takayanagi, ``Boundary states for the rolling
  D-branes in NS5 background,'' {\em JHEP} {\bf 07} (2004) 020,
\href{http://www.arXiv.org/abs/hep-th/0406173}{{\tt hep-th/0406173}}.

\bibitem{Ooguri:1995wj}
H.~Ooguri and C.~Vafa, ``Two-Dimensional Black Hole and Singularities of CY
  Manifolds,'' {\em Nucl. Phys.} {\bf B463} (1996) 55--72,
\href{http://www.arXiv.org/abs/hep-th/9511164}{{\tt hep-th/9511164}}.

\bibitem{Mukhi:1993zb}
S.~Mukhi and C.~Vafa, ``Two-dimensional black hole as a topological coset model
  of c = 1 string theory,'' {\em Nucl. Phys.} {\bf B407} (1993) 667--705,
\href{http://www.arXiv.org/abs/hep-th/9301083}{{\tt hep-th/9301083}}.

\bibitem{Ghoshal:1995wm}
D.~Ghoshal and C.~Vafa, ``C = 1 string as the topological theory of the
  conifold,'' {\em Nucl. Phys.} {\bf B453} (1995) 121--128,
\href{http://www.arXiv.org/abs/hep-th/9506122}{{\tt hep-th/9506122}}.

\bibitem{Ghoshal:1993qt}
D.~Ghoshal and S.~Mukhi, ``Topological Landau-Ginzburg model of two-dimensional
  string theory,'' {\em Nucl. Phys.} {\bf B425} (1994) 173--190,
\href{http://www.arXiv.org/abs/hep-th/9312189}{{\tt hep-th/9312189}}.

\bibitem{Hikida:2004mp}
Y.~Hikida and T.~Takayanagi, ``On Solvable Time-Dependent Model and Rolling
  Closed String Tachyon,''
\href{http://www.arXiv.org/abs/hep-th/0408124}{{\tt hep-th/0408124}}.

\bibitem{Kazakov:2000pm}
V.~Kazakov, I.~K. Kostov, and D.~Kutasov, ``A matrix model for the
  two-dimensional black hole,'' {\em Nucl. Phys.} {\bf B622} (2002) 141--188,
\href{http://www.arXiv.org/abs/hep-th/0101011}{{\tt hep-th/0101011}}.

\bibitem{Giveon:2003wn}
A.~Giveon, A.~Konechny, A.~Pakman, and A.~Sever, ``Type 0 strings in a 2-d
  black hole,'' {\em JHEP} {\bf 10} (2003) 025,
\href{http://www.arXiv.org/abs/hep-th/0309056}{{\tt hep-th/0309056}}.

\bibitem{Carlisle:2004jn}
J.~E. Carlisle and C.~V. Johnson, ``Unoriented minimal type 0 strings,''
\href{http://www.arXiv.org/abs/hep-th/0408159}{{\tt hep-th/0408159}}.

\bibitem{Klebanov:2003km}
I.~R. Klebanov, J.~Maldacena, and N.~Seiberg, ``D-brane decay in
  two-dimensional string theory,'' {\em JHEP} {\bf 07} (2003) 045,
\href{http://www.arXiv.org/abs/hep-th/0305159}{{\tt hep-th/0305159}}.

\bibitem{Govindarajan:2003vv}
S.~Govindarajan and J.~Majumder, ``Orientifolds of type IIA strings on
  Calabi-Yau manifolds,'' {\em Pramana} {\bf 62} (2004) 711--716,
\href{http://www.arXiv.org/abs/hep-th/0305108}{{\tt hep-th/0305108}}.

\bibitem{Govindarajan:2003vp}
S.~Govindarajan and J.~Majumder, ``Crosscaps in Gepner models and type IIA
  orientifolds,'' {\em JHEP} {\bf 02} (2004) 026,
\href{http://www.arXiv.org/abs/hep-th/0306257}{{\tt hep-th/0306257}}.

\bibitem{Blumenhagen:2003su}
R.~Blumenhagen, ``Supersymmetric orientifolds of Gepner models,'' {\em JHEP}
  {\bf 11} (2003) 055,
\href{http://www.arXiv.org/abs/hep-th/0310244}{{\tt hep-th/0310244}}.

\bibitem{Aldazabal:2003ub}
G.~Aldazabal, E.~C. Andres, M.~Leston, and C.~Nunez, ``Type IIB orientifolds on
  Gepner points,'' {\em JHEP} {\bf 09} (2003) 067,
\href{http://www.arXiv.org/abs/hep-th/0307183}{{\tt hep-th/0307183}}.

\bibitem{Brunner:2004zd}
I.~Brunner, K.~Hori, K.~Hosomichi, and J.~Walcher, ``Orientifolds of Gepner
  models,''
\href{http://www.arXiv.org/abs/hep-th/0401137}{{\tt hep-th/0401137}}.

\bibitem{Giveon:1998sr}
A.~Giveon and D.~Kutasov, ``Brane dynamics and gauge theory,'' {\em Rev. Mod.
  Phys.} {\bf 71} (1999) 983--1084,
\href{http://www.arXiv.org/abs/hep-th/9802067}{{\tt hep-th/9802067}}.

\bibitem{Kutasov:2004dj}
D.~Kutasov, ``D-brane dynamics near NS5-branes,''
\href{http://www.arXiv.org/abs/hep-th/0405058}{{\tt hep-th/0405058}}.

\bibitem{Angelantonj:2002ct}
C.~Angelantonj and A.~Sagnotti, ``Open strings,'' {\em Phys. Rept.} {\bf 371}
  (2002) 1--150,
\href{http://www.arXiv.org/abs/hep-th/0204089}{{\tt hep-th/0204089}}.

\end{thebibliography}\endgroup

\end{document}